\documentclass[]{aa} 
\usepackage{graphicx}

\begin{document}
   \title{Star formation history of CMa R1}

  \subtitle{I. Wide-field X-ray study of the young stellar population\thanks{Based in part on observations obtained at the Gemini Observatory, 
which is operated by the
Association of Universities for Research in Astronomy, Inc., under a cooperative agreement
with the NSF on behalf of the Gemini partnership: the National Science Foundation (United
States), the Science and Technology Facilities Council (United Kingdom), the
National Research Council (Canada), CONICYT (Chile), the Australian Research Council
(Australia), Minist\'erio da Ci\^encia e Tecnologia (Brazil) 
and Ministerio de Ciencia, Tecnolog\'ia e Innovaci\'on Productiva  (Argentina)
}}

\author{J. Gregorio-Hetem
          \inst{1}
          \and
          T. Montmerle\inst{2}
\and
C. V. Rodrigues\inst{3}
\and
E. Marciotto\inst{1}
\and
T. Preibisch\inst{4}
\and
H. Zinnecker\inst{5}
          }

   \offprints{J. Gregorio-Hetem}

   \institute{Universidade de S\~ao Paulo,
              IAG, Departamento de Astronomia, Brazil,
              \email{jane@astro.iag.usp.br}
         \and
Laboratoire d'Astrophysique de Grenoble, Universit\'e Joseph Fourier-CNRS, France             
             %\email{montmerle@obs.ujf-grenoble.fr}      
\and
Instituto Nacional de Pesquisas Espaciais, Divis\~ao de Astrof\'isica, S\~ao Jos\'e dos Campos, SP, Brazil.
%\email{claudiavr@das.inpe.br}
\and
Universit\"ats-Sternwarte M\"unchen,
  Scheinerstr.~1, D-81679 M\"unchen, Germany
\and
Astrophysikaliches Institut Potsdam
}

 %  \date{Received September 15, 1996; accepted March 16, 1997}

% \abstract{}{}{}{}{} 
% 5 {} token are mandatory
 
  \abstract
{}
%{\sl Aims}. 
{The CMa R1 star-forming region contains several compact clusters as well as many young early-B stars. It is associated with a well-known 
bright rimmed nebula, the nature of which is unclear (fossil HII region or supernova remnant). To help elucidate the nature of the nebula, 
our goal was to reconstruct the star-formation history of the CMa R1 region, including the previously unknown older, fainter low-mass 
stellar population, using X-rays.}
%{\sl Methods}. 
{We analyzed images obtained with the {{\it ROSAT}} satellite, covering $\sim 5$ sq. deg. Complementary VRI photometry was performed with 
the {{\it Gemini}} South telescope. Colour-magnitude and colour-colour diagrams were used in conjunction with pre-main sequence evolutionary 
tracks to derive the masses and ages of the X-ray sources.}
%{\sl Results}. 
{The {{\it ROSAT}} images show two distinct clusters. One is associated with the known optical clusters near Z CMa, to which $\sim 40$ members  
are added. The other, which we name the ``GU CMa'' cluster, is new, and contains $\sim 60$ members. The {{\it ROSAT}} sources are young 
stars with masses down to $M_\star \sim 0.5 M_\odot$, and ages up to 10 Myr. The mass functions of the two clusters are similar, but the 
GU CMa cluster is older than the cluster around Z CMa by at least a few Myr. Also, the GU CMa cluster is away from any molecular cloud, 
implying that star formation must have ceased; on the contrary (as already known), star formation is very active in the Z CMa region.}
%{\sl Conclusions}.
{}
%{The CMa R1 region has undergone at least two distinct star formation episodes. Only the current Z CMa star formation activity seems %related to the CMa R1 nebula and possibly triggered by it. Star formation in GU CMa precedes it by several Myr, and is therefore %unrelated to the nebula.}


\authorrunning{Gregorio-Hetem et al.}
\titlerunning{Star formation history of CMa R1}

   \maketitle
%
%________________________________________________________________
%%========================================Sect. 1

\section{Introduction}
CMa R1 is an association of bright stars and clusters distributed around and in the vicinity of the long (200') arc-shaped 
ionized reflection nebula Sh2-296  (Sharpless 1959: l=224.6$^o$, b=-2$^o$), located at a distance $d \sim 1$ kpc (Shevchenko et al. 1999, 
Kaltcheva \& Hilditch 2000). In the absence of conspicuous exciting early-type stars inside the arc, the source of ionization is still being debated. 
The CMa R1  nebulae are found within the boundaries of the OB1 association, which approximately are: 
$222^o < l < 226^o$ and $-3.4^o < b < +0.7^o$ (Ruprecht 1966). Clari\'a (1974a,b) studied the space distribution of O and B stars,
based on UBV photometry obtained for 247 stars.  The estimated E(B-V) colour excess confirms the existence of a group of young OB stars
together with excited gas and obscuring matter, which belong to the CMa OB1 association.  Clari\'a used these data to derive an age of 3 ~Myr.

In the early years of the `propagating star formation' models (Elmegreen \& Lada 1977), Herbst \& Assousa (1977) suggested
that the ``CMa R1 ring" (Sh2-296) could be an old  supernova remnant (SNR), which was inducing star formation in CMa R1. 
Indeed, linear polarization observations are consistent with a model of compression by a supernova shock (Vrba, Baierlein \& Herbst 1987).
Alternatively, Reynolds \& Ogden (1978) proposed that the nebula and star formation were induced by strong stellar winds, or by an evolving, 
`fossil' HII region, as also suggested by Blitz (1980) and Pyatunina \& Taraskin (1986).
However, because they are found to be in pressure equilibrium with the parent molecular cloud once they have opened a cavity 
(see the example of M17, Townsley et al. 2003, or Orion, G\"udel et al. 2008), stellar winds cannot compress the surrounding medium and induce star formation. 
On the other hand, we now know many examples of HII ``bubbles" apparently triggering star formation near their edges 
(e.g., Deharveng et al. 2008, Zavagno et al. 2007), but these bubbles have well-defined exciting stars.

\hspace{-2cm}
%%-------------------------------------------------------Figure 1----------- IRAS-DSS
\begin{figure*}[ht]
\begin{center}
%\vspace{-5cm}
%\includegraphics[bb=62 69 506 550,height=18.5cm,angle=270,clip]{12140f01.jpg}
\includegraphics[height=18.5cm,angle=270,clip]{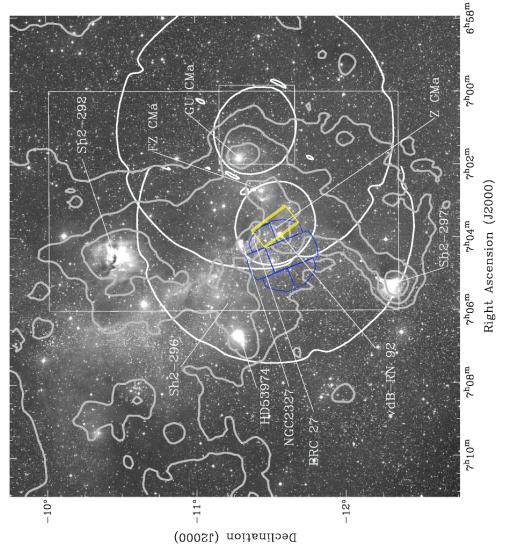}
\caption{Finding chart of the CMa OB1/R1 region. Far-infrared IRAS-ISIS contours superimposed on a 3 $\times$ 3 square degrees optical image 
extracted from the Digitized Sky Survey. The grey contours correspond to data obtained at 100 $\mu$m, tracing the dust component.
The field-of-view of X-ray observations are indicated for: {\it ROSAT} (white lines); {\it XMM-Newton} (blue) and Chandra (yellow).
The large rectangular white box shows the area covered by the optical survey of OB stars by Shevchenko et al. (1999), while the small square box 
including GU CMa corresponds to the area where {\it Gemini} observations of unresolved X-ray sources were carried out. Objects discussed in the
text are labelled by their names.
North is up and East is to the left.}
\end{center}
\end{figure*}

%%--------------------------------------------------------EndFig.1-----------------------------

To this day, the nature of the Sh2-296 nebula remains a mystery, yet   we see that star formation is going on in the vicinity of the nebula.  
The location of the Sh2-296 arc and of other visible sites of star  formation, in particular the associated nebulae Sh2-292 (IC2177) and  
Sh2-297 (HD53623),  with respect to the cold dust distribution, are shown in Figure 1, in the form of {\sl IRAS} 100 $\mu$m contours, 
superimposed on a DSS optical (R) image. It is clear that the sites of star formation correspond to the largest dust column densities, 
mainly distributed along the edge of the arc.

The molecular gas around CMa R1 has also been studied in the course of millimeter surveys of the third galactic quadrant. The $^{12}$CO($1 \rightarrow 0$) 
Columbia survey of May et al. (1988), with a spatial resolution of $0.5^o$, reveals the extent, down to low densities, of the molecular clouds in the CMa R1 region. 
Taking into account the relatively low spatial resolution of the survey, the $^{12}$CO contours match the {\sl IRAS} 100$\mu$m maps. The densest regions are 
outlined in the  $^{13}$CO($1 \rightarrow 0$) Nanten survey of Kim et al. (2004), which has a better spatial resolution (8.8' spacing). 
These regions are the ``backbone" of molecular clouds, where the earliest stages of star formation can be found. Here they are not so well correlated with 
{\sl IRAS} maps, but, as expected, underlie the other nebulae. These CO maps, with which we will compare the young stars distribution, give precious clues about 
the star formation history of the region. (See below, Sect. 6 and Fig.12, for further discussion.)

%%-------------------------------------------------------Figure 2----------- ROSAT fields
\begin{figure*}[!t]
\includegraphics[height=18.5cm,angle=270,clip]{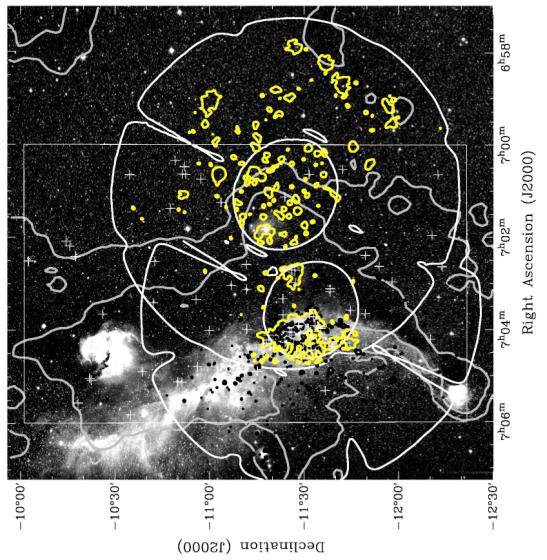}
\caption{{\it ROSAT} PSPC X-ray contours and fields-of-view superimposed on a digitized POSS(R) image of the CMa R1 region. 
Grey contours show the same  IRAS-ISIS data at 100 $\mu$m as Fig.1. The observations of Field~1 are indicated by yellow contours and of Field~2 by black contours.
The white crosses indicate the association members identified in the optical survey by Shevchenko et al. (1999), within the area outlined by a white rectangular box.
}
\end{figure*}
%%--------------------------------------------------------EndFig.2-----------------------------

Earlier works already contain evolutionary information about the stellar population in the CMa R1 region. 
A survey of early-type stars was done by Shevchenko et al. (1999) covering a 4 sq. deg. rectangular area that overlaps the 
Sh2-296 arc, revealing 88 members of the CMa R1 association (74 B stars, no O star) with ages between 8 and 0.5 Myr, apparently 
suggesting an extended period of star formation activity that is still ongoing.

Shevchenko et al. (1999) also point 
out that two other bright B stars, GU CMa and FZ CMa, seem to be older than the association and may not have been formed in the same star formation episode.
Other early B stars associated with Sh2-296 include Z CMa, a double system (Koresko et al. 1991, Haas et al 1992) containing a FU Orionis
object (Hartmann et al. 1989) and a Herbig Be star, with a common age of about 0.3 Myr (van den Ancker et al. 2004). 

In addition, three embedded stellar clusters, NGC2327, BRC 27, and VdB-RN92, are found along the outer rim of Sh2-296, in the vicinity of  
Z CMa. These clusters were  studied by Soares \& Bica (2002, 2003), who used JHK data from 2MASS (``The Two Micron All Sky Survey", 
Skrutskie et al. 2006). Based on colour-magnitude diagrams, they estimated ages of 1.5 Myr (NGC2327 and BRC27)
and 5-7 Myr (VdB-RN92), spanning the same age range as the sample of Shevchenko et al. (1999).

We note, however, that such surveys, by construction, miss the most numerous, fainter low-mass ($< 2 M_\odot$) star population, which may have a different spatial 
distribution, as well as different ages. 
The most efficient method to find young stars including low-mass ones is to use X-rays, the lower-mass limit being only a 
question of sensitivity for a given instrument, i.e., of exposure time and distance to the star-forming region. The reasons are twofold: $(i)$ young stars of all masses 
produce X-rays (except perhaps A stars, see, e.g., Stelzer et al. 2005), independently of their circumstellar environment : massive stars via shocks in their winds, 
with a typical X-ray-to-bolometric luminosity ratio $L_X/L_{bol} \sim 10^{-7}$, low-mass stars via their magnetic activity, with $L_X/L_{bol} \sim 10^{-5}-10^{-3}$; $(ii)$ X-ray 
satellites have a wide field-of-view, covering up to several square degrees in one exposure. (For reviews, see Feigelson \& Montmerle 1999, Favata \& Micela 
2003, G\"udel 2004.)

By having access in this way to a much more complete census of the  stellar population, down to low-mass stars, we will show in this paper how we can significantly improve 
our knowledge of the evolution of star formation in the area, both in time (from colour-colour diagrams) and in space (wide field-of-view of X-ray telescopes), 
and  thus investigate whether Sh2-296, whatever its nature, may have triggered star formation in the CMa R1 region and its vicinity.

More precisely, in order  to obtain a stellar sample covering as large a range of masses and ages as possible, we have analyzed archival X-ray data from 
the {\sl PSPC} detector aboard the {\sl ROSAT} satellite, because of its wide field of view ($1^{o}$ radius, or an area of $\sim 3$ sq. deg.).

We will proceed along the following steps:  $(i)$ identify the X-ray sources, by comparison with data from optical and IR surveys, as well as from dedicated {\sl Gemini} 
VRI observations; $(ii)$ put the counterparts on appropriate colour-magnitude diagrams and deduce their evolutionary status (mass and age) using PMS evolution 
models; $(iii)$  study the spatial distribution of the X-ray sources and relate it to the mass distribution in its vicinity. In future papers, we will `zoom in' towards more 
restricted regions in the Z CMa region, using {\sl XMM-Newton/EPIC}  
(circle of $15'$ radius) and {\sl Chandra/ACIS} (square FOV, $17'  
\times 17'$) .

Although we have used the XMM-Newton and Chandra source positions,  
when available, to validate the ROSAT positions in Field 2, we defer a more  
detailed analysis of these fields (source luminosities, spectra, etc.)  
to a future study  (see a preliminary results by Rojas et al. 2005)  
for several reasons: (i) as shown in Fig.1, the XMM-Newton and Chandra  
observations cover a much smaller area than the ROSAT fields; (ii) the  
50 ksec XMM-Newton total exposure is heavily affected by solar flares,  
and thus requires a detailed treatment to recover as much information  
as possible; (iii) in contrast, the Chandra observations are much more  
sensitive (total exposure of 80 ksec in two observations) and thus  
contain many sources which ROSAT is unable to detect in Field 2: obviously this  
would distort the comparison between the two ROSAT fields.

This paper is therefore organized as follows. Sect. 2 describes  
archival {\sl ROSAT/PSPC} observations of the CMa R1 region and the  
general characteristics of the X-ray data. Stellar identifications based on optical and near-IR counterparts searched in published
catalogues are described in Sect. 3. The {\it  Gemini} 
observations, which were obtained specifically to search for additional counterparts in the case of unresolved X-ray sources, and their analysis, are
presented in Sect. 4. Ages and masses of the {\it ROSAT} sources are derived from colour-magnitude diagrams in Sect. 5.

The last section gives a summary of the results, followed by a discussion and conclusions. 
Appendix A gives
details on the {\sl ROSAT} data analysis, and Appendix B gives
complementary results of the {\it Gemini} data analysis.

%========================================Section 2

\section{Search for young stars based on X-ray data}

The power of X-ray observations to discover large samples of PMS stars has been illustrated in a variety of star-forming regions. In particular, following up 
on pioneering {\sl ROSAT} observations, intermediate to distant clouds have been covered 
by {\sl Chandra}: the Orion nebula (Garmire et al. 2000; Feigelson et al. 2002, 2003; Flaccomio et al. 2003), part of the Rosette 
nebula and molecular cloud, together with M17 (Townsley et al. 2003; Wang et al. 2008); Mon R2 (Kohno et al. 
2002); RCW38 (Wolk et al. 2002); M16 (Linsky et al. 2007), 
and by {\sl XMM-Newton}: M8 (Rauw et al. 2002);  Carina (Albacete Colombo et al. 2003); Vela OB2 (Jeffries et al. 2009); 
Orion (L\'opez-Santiago \& Caballero 2008); and others.
Up to thousands of point
X-ray sources per cloud are detected, almost all identified with young stars, down to the 
brown dwarf regime. More than 1600 X-ray sources were detected in the Orion 
Nebula Cluster during an exceptionally deep survey ($\sim 1$ Msec), the ``Chandra Orion Ultradeep Project'' 
(COUP: Getman et al. 2005a, 2005b), and several hundred others have been found in the recent ``{\sl XMM-Newton} 
Extended Survey of Taurus" (XEST: G\"udel et al. 2007).

However, in spite of the much improved sensitivity and angular resolution of {\sl Chandra} and {\sl XMM-Newton} over {\sl ROSAT},
there are spatial limitations to these observations. To map a wide area (several square degrees), a mosaic is necessary: 
for instance, 5  {\sl Chandra} fields for the Rosette nebula and associated molecular cloud (Wang et al. 2008), 28 {\sl XMM-Newton} 
fields to map the densest molecular clouds of Taurus (G\"udel et al. 2007). In the case of Sh2-296, two overlapping archival 
{\sl ROSAT/PSPC} fields exist (see Fig. 1), covering $\sim 5$ sq. deg. in total, which we are analyzing in this paper.
The present {\sl ROSAT} work on CMa R1 is related to a similar study based on observations of two 
other giant molecular clouds, Monoceros and Rosette (Gregorio-Hetem et al. 1998). The {\it ROSAT} images
provided the identification and characterization of several dozen sources,making up the bright end of the low- 
to in\-ter\-me\-diate-mass young star distribution in the associated clusters. 
In particular, a useful correlation is found between the intrinsic X-ray luminosity of the sources and their optical and 
near-IR magnitudes.
This correlation was similar to the one found previously
by Feigelson et al. (1993) and Casanova et al. (1995) from {\sl ROSAT} studies of the nearby  
Chamaeleon I and $\rho$ Ophiuchi clouds, respectively, suggesting a general validity which we will use in the  
present paper.  The main interest in this correlation is that it is weakly dependent on extinction, which is essential
for the study of young stars, because of the comparable absorption cross-sections in the near-IR and keV ranges
(Ryter 1996).

The two {\sl ROSAT} observations we analyze are the following. By  order of increasing RA, the first field (``Field 1" hereafter),
HEASARC ID RP201011 (PI J.G.H.), pointing axis $\alpha_{J2000}$ =  $7^h00^m$,  $\delta_{J2000}$ = $ -11^o 30'$, has an exposure 
of 19.7  ksec.
(For preliminary results, see Gregorio-Hetem, Montmerle, \& Marciotto  
2003.) ``Field 2", HEASARC ID RP201277 (PI H.Z.), pointing axis
$\alpha_{J2000}$ = $7^h04^m$,  $\delta_{J2000}$ = $ -11^o 33'$, has a  
much shorter exposure of 4.6 ksec, and has been partially published as
part of a survey of Herbig AeBe stars, here Z CMa (Zinnecker \&  
Preibisch 1994). In such exposures, the number of X-ray counts per  
source is
in general too small to allow deriving a spectrum, or even a good  
hardness ratio. Therefore, the intrinsic X-ray luminosity of the  
sources, known to
be PMS stars in this case, is based on the count rate, assuming an average plasma temperature (here $kT_X \sim 1$ keV), and is corrected for 
extinction using the column density given by $N_H = 2.1 \times 10^{21} A_V {\rm cm}^{-2}$ (e.g., Vuong et al. 2003), $A_V$ being determined by 
independent methods (see below). A typical value of the approximate X-
ray luminosity $\tilde{L}_X$ for $A_V \sim 1$ is based on the  {\sl  
ROSAT}
count-to-flux ratio 1 ct~ksec$^{-1}$ = $9 \times 10^{-15} {\rm  
erg~s}^{-1}~{\rm cm}^{-2}$ (see details in Appendix A).  The  
corresponding lower limits to the
stellar X-ray  luminosities are $\tilde{L}_{X,min} \sim 6 \times  
10^{29} {\rm erg~s}^{-1}$ for Field 1, and $\tilde{L}_{X,min} \sim 1  
\times 10^{30} {\rm erg~s}^{-1} $
for Field 2.

Figure 2 is a composite image of Fields 1 and 2, superimposed on a digitized POSS(R) image. The analysis of Field 1 (west of Sh2-296; exposure $\sim 20$ ksec) 
reveals 61 X-ray sources, the majority of them (47)
 located in the more sensitive central part of the {\sl ROSAT} field. Field 2 (overlapping Sh2-296; exposure $\sim 5$ ksec) contains 37 more sources, 
which appear much more concentrated in the central region; the outer regions are probably not sufficiently exposed to reveal additional sources. The clustering of 
Field 2 sources does appear in Field 1, in the form of an extended area of emission,
but, being close to the edge of the {\sl PSPC} FOV, the sources are not resolved individually and thus are 
not included in the list of Field 1 sources. In contrast, these sources are resolved in Field 2.
Five  sources are detected in the overlapping area between the two fields.
In total, 98 {\it ROSAT} sources were detected  in these fields. Table A.1 lists 56 distinct sources belonging to Field 1, and 37 belonging to Field 2, and 5 
more being detected in both fields.
Most of the sources have at least one optical counterpart.  Based on their count rates and the estimate of their bolometric 
luminosities (see next section), we find log$(\tilde{L}_X/L_{bol})$ lying in the -6 to -4 range, typical of low-mass young stars. 

What is equally remarkable is the {\it absence} of sources in some areas. In Field 2, the near-absence of sources elsewhere than in the most sensitive 
central {\sl PSPC} ring is very likely an observational bias resulting from the short exposure: 
except perhaps for the brightest, several of the sources visible in Field 1 would not be detected in Field 2 (detection limit: log$\tilde{L}_X = 30$). 
In contrast, in Field 1 the absence of sources outside the detected X-ray clustering is very significant. In particular, in the overlapping area between 
Field 1 and the survey area of Shevchenko et al. (1999) (see Fig. 2), no star is detected North and South of GU CMa ($\alpha \approx$ 7h02m). It is 
therefore clear that {\it the clustering of X-ray sources visible in Field 1 corresponds to a physical cluster}, spatially distinct from the B stars identified 
by Shevchenko et al. (1999). GU CMa certainly belongs to this new cluster, and it is probable that FZ CMa does too, confirming the suspicion of 
Shevchenko et al. (1999) that these stars may have originated in a different star formation episode. We will address this question in detail in the 
next sections, adopting the name ``GU CMa" for the new cluster.

Appendix A describes the X-ray observations and Table A1 lists the
98 sources detected in the {\it ROSAT} fields, which we name CMaX-(number) for convenience. 
Two methods were adopted to look for counterparts of the CMaX sources: (i) using
available catalogues (described in Sect. 3), and (ii) in case no counterpart was found, using photometric data from follow-up {\it Gemini} observations (see Sect. 4).

%%----------------------------------------------Figure 3 near-IR colour-colour
\begin{figure}[ht]
\includegraphics[height=8cm,angle=270]{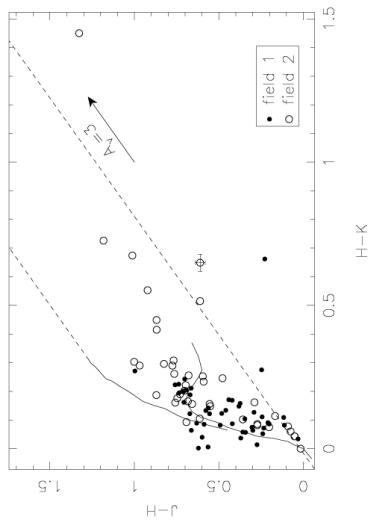}
\caption{Diagram J-H vs. H-K of resolved CMaX sources
detected in {\it ROSAT} fields 1 (filled circles)  and 2 (open circles).
Full lines indicate the main sequence and red giant branch, and 
dashed lines are used to show the direction of the interstellar reddening vector. 
Error bars are shown for a representative CMaX source.}

\end{figure}
%%------------------------------------------end of Fig. 3

%%------------------------------Figure 4 central part of ROSAT field
\begin{figure*}[ht]
%\sidecaption
%\vspace{-1cm}
%\includegraphics[bb=3 3 641 588,width=15cm,angle=0]{12140f04.jpg}
\includegraphics[width=15cm,angle=0]{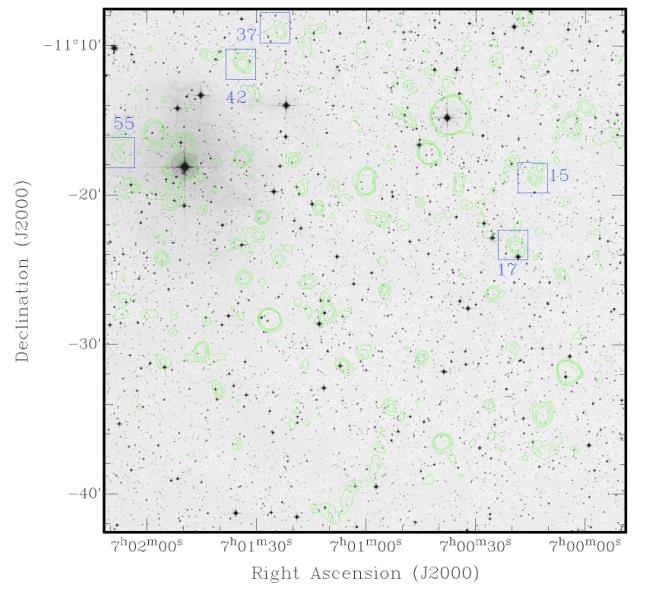}
\caption{X-ray sources (green contours) and observed {\it Gemini}
fields (blue squares, labelled by the corresponding CMaX numbers), superimposed on the POSS(R) image. The X-ray
contours were extracted from the {\it ROSAT} image using a block bin factor of 30 (see 
Sect. 4.4), smoothed using a Gaussian PSF with $\sigma$ = 2 pixels.
}
\end{figure*}

%%========================================Sect. 3
\section{Stellar identifications}

From the available images and catalogues listed below, we found
at least one optical and/or infrared counterpart for 91\% of the 
{\it ROSAT} sources, 13 of them having known spectral types; the most massive are five B2-3 type stars.
Nine X-ray sources coincide with the emission line stars surveyed by Clari\'a (1974) and/or 
Schevchenko el al. (1999); their identifications from these catalogues are denoted by
``C" and ``S" respectively in Table A1. We note that Z CMa, FZ CMa and GU CMa are detected respectively as CMaX-63, -57, and -48.
In addition, nine {\it ROSAT} sources coincide with the Soares \& Bica clusters: two in BRC27 (CMaX-74, -75), three in NGC2327
(CMaX-76, -77, -81), and four in vdB-RN92 (CMaX-66, -68, -72, -73). These identifications are also noted in Table A1.

Optical counterparts for the CMaX sources of Field 1 were first identified on a POSS(R) plate kindly digitized for us by  the
MAMA\footnote {Machine
 Automatique \`a Mesurer pour l'Astronomie, Observatoire de Paris; now discontinued operation.} device.
However the corresponding MAMA data are not available for Field 2, so for consistency we use the USNO Catalogue for the identification and R magnitudes of the optical counterparts in both fields.
The 2MASS catalogue was also used to obtain the IR photometry of the stellar counterparts in the JHK bands. 
The adopted positional accuracy of the optical counterparts is based on the brightest star located inside the {\it ROSAT} error circle. 
Table A1 gives these errors, which are less than 10 arcsec for
sources in the central areas, and 30 arcsec at the edge of the fields.

We have also used the
 {\it XMM-Newton and Chandra} data, when available, to check the consistency of the
positional accuracy of the optical counterparts to the CMaX sources located in the center area of Field 2, which corresponds to the best {\it ROSAT} resolution. Fourteen CMaX sources (listed in Table A3) have been detected by {\it XMM-Newton and Chandra} in this area, confirming the identifications based on {\it ROSAT} positions.

Figure 3 shows the J-H {\it vs.} H-K diagram obtained for the near-IR counterparts. This diagram 
indicates that most CMaX sources do not suffer high extinction. Those above the main 
sequence are located in the direction of the reddening lines, allowing a reliable extinction correction. 
As explained below (Sect. 4.3), the reddening was estimated from the visual extinction derived 
from star counts, and checked using two control regions containing field stars.
However, four objects present a real H-K excess (appearing to the right side of the dashed line in Fig. 3), which is an indicator of the presence of circumstellar matter. Among them, we  
find the well-known Herbig star Z CMa, which has H-K $\sim$ 1.5 mag.
%Some of the X-ray fields contain sources having no counterpart, inside the position error circle, identified  in the published 
%catalogues or images. 
There are also some of the X-ray fields for which we could not find any counterpart. 
In that case, the position of the
 X-ray centroid is indicated between parentheses in Table A1.
Since these sources have X-ray luminosities in the range log(\~L$_X$) $\sim$ 29.4-30.4, one would expect a counterpart
 brighter than $R \sim 16$, based on the abovementioned (L$_X$,$M_R$) correlation holding for young low-mass stars (further 
discussed in Sect. 4.4).  In a search for counterparts fainter than the POSS(R) plate or USNO limit (R $\sim$19 mag), we have obtained deeper observations 
in the V, R and I bands with the {\it Gemini} South telescope. Combining them with the 2MASS catalogue, we obtain altogether VRIJHK photometry for the detected counterparts.
The search for these additional, faint counterparts to the unresolved {\it ROSAT} sources, which we name the ``{\it Gemini} candidates", is discussed  in 
the next Section.

%%%%%%%%%%%%%%%%%%%%%%%%%%%%%%%%%Start Sect.4%%%%%%%%%

 \section{Search for faint counterparts}
%========================================Sect. 4.1
\subsection{{\rm Gemini} VRI images and optical magnitudes}

%Five fields were observed with {\it Gemini}, searching for optical counterparts to ``empty" X-ray contours. 

Five fields were observed with {\it Gemini}, searching for optical counterparts to X-ray contours for which the centre is ``empty".
   
Their positions, all within the {\it ROSAT} Field 1, are shown in Figure 4. We use the correlations between X-ray luminosities and 
absolute magnitudes (log(\~L$_X$) vs. $M_{R}$ and log(\~L$_X$) vs. $M_{J}$), obtained for young low-mass stars  and displayed 
in Figs. A1 and A2. These correlations show 1$\sigma$ deviations of 1.4 mag in R band and 1.2 mag in J. We estimated the expected apparent R and J
 magnitudes of these counterparts, using a distance modulus 
of 10 mag, with the result that these magnitudes are well above the POSS(R) and 2MASS limits. Yet the counterparts to such bright X-ray sources do not have corresponding optical magnitudes: CMaX15, for example, should be associated with an object having R=(13.2 $\pm$ 1.4)  mag, but the brightest object within the X-ray contours has R=17.1. These 
numbers are consistent with the published USNO accuracy ($\sigma$ = $\pm 0.3$~mag). 
The L$_X$/$M_{R}$ 
ratio in this case is 3.9 orders of magnitude higher than the average value observed in T Tauri stars, i.e., 2.8$\sigma_R$ away
from the correlation. This fact motivated us to  search for multiple faint counterparts that could correspond to unresolved X-ray emitters.

%%---------------Fig. 5 --------------------------Photometric Calibration
\begin{figure}[!t]
%\vspace{6.5cm}
\centering
\includegraphics[height=7cm,angle=0]{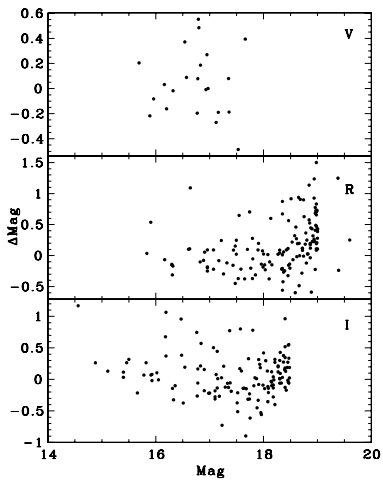}
\caption{Difference between our calibrated and catalogue magnitudes 
as a function of catalogue magnitude 
for objects in our {\it Gemini} fields. From top to bottom, the panels 
give the V, R, and I measurements.
} 
\end{figure}
%%--------------- End Fig. 5

Table 1 gives coordinates of the centroid of the target {\it ROSAT} sources, their X-ray properties, and their
expected R and J apparent magnitudes.
It also gives the number of candidate counterparts for each field, as explained in Sect. 4.2, and their extinction, detailed in Sec. 4.3.

The observations were done with the Acquisition Camera at the 8.1m {\it Gemini} South telescope, which provides 
$\sim$ 2' x 2' images with a resolution of 0.12 
arcsec/pixel. We used the V 
($\lambda_c$=540nm), R ($\lambda_c$=640nm) and I ($\lambda_c$=790nm) filters, 
 with total exposure times of 360~s for the V images, and 300~s for R 
and I. The fields containing sources CMaX-15, 37, and 55 were 
observed in 2001, October (Program ID: GS-2001B-Q41), while CMaX-17 and 42 were 
observed in 2003, February (Program ID:GS-2003A-Q06). This last run had weather conditions better 
than those of the first run. 
The usual data reduction has been made using bias, flat-field, and dark corrected
images provided by the {\it Gemini} Observatory.

The photometric calibration of the images was performed using common 
objects in the USNO B1.0 catalogue (Monet et al. 2003) for the I filter 
and in the NOMAD catalogue (http://www.navy.mil/nomad.html) for the V and R filters. 
The number of objects used in the calibration is typically 20 
in R and I, and 5 in V. The brighter objects, which 
are saturated in the {\it Gemini} images, have not been 
used in the calibration procedure. We have transformed the USNO R magnitude into the Landolt system 
using the expression given by Kidger (2003). No colour 
correction was applied.

Figure 5 shows a comparison 
between the magnitudes in the NOMAD and USNO catalogues and those 
obtained using our calibration (Monet et al. 2003).  We have measured the magnitudes of over a 
hundred stars in each field. The RMSs of 
these differences are: 0.26 (V), 0.44 (R) and 0.35 (I). These 
numbers are consistent with the published USNO accuracy ($\pm 0.3$~mag).

%%-------------------------------------------------------Figure 6 ------------------ Gemini and  X-rays
\begin{figure}[ht]
\includegraphics[height=8.7cm,angle=270,clip]{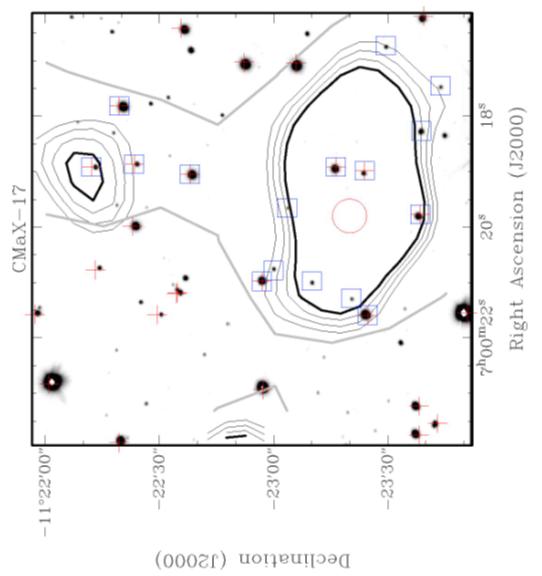}
\includegraphics[height=8.0cm,angle=0]{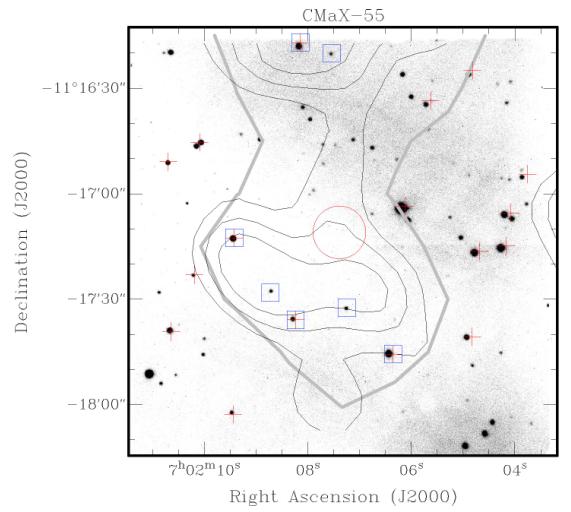}
\caption{Optical image obtained by {\it Gemini} (I band) of CMaX-17 (top) and CMaX-55 (bottom), two examples 
of ``extended'' CMaX sources. Black contours show the X-ray map having enhanced resolution  (5 arcsec/pixel),
while the nominal resolution (30 arcsec/pixel) is illustrated by the grey line. 
A red circle indicates the {\it ROSAT} nominal position error; red crosses show the 2MASS sources present in the {\it Gemini} field,
while blue squares show the putative candidates.} 
\end{figure}
%%-------------------------------------------------------End Figure 6

%======================================== Sect. 4.3 

\subsection{Finding Candidate Counterparts}

The nominal angular resolution of the standard {\it ROSAT} image shown in Figure 2 is 30 arcsec/pixel. To bring out spatial details in X-rays, we extracted images having an enhanced resolution (5 arcsec/pixel) for each of the {\it Gemini} fields, by adopting a blocking factor of 10. 
In this way we can map X-ray contours, including sub-structures, in order to look for unresolved optical candidates that may contribute to the X-ray emission.

%%-----------------------------------------------  New Table 1 ----------------------
\begin{table*}[ht] 
\caption{ List of CMaX sources observed with {\it Gemini}.} 
\begin{center}
{\scriptsize
\begin{tabular}{l|cccccc|cc|cc|c}
\hline
& 
\multicolumn{6}{c}{X-ray source} &
\multicolumn{2}{|c}{Expected$^a$} &
\multicolumn{2}{|c|}{Candidates$^b$} & \\
%\hline
CMaX& R.A. & Dec. & Err & Cnts/ks & S/N & log$L_X$ &J & R & 
Optical & Near-IR&  $A_{V}$ \\
& J2000 & J2000 &(") & & & (erg/s) & (mag) & (mag) & & & (mag)  \\
\hline
15& 07:00:15 & -11:18:51 & 10 &1.9 & 4.8& 30.39 & 11.8 & 13.3 &
12& 6 &  1.23$\pm$0.03\\
17N& 07:00:19 & -11:22:10 & 9 & 0.38 & 1.7 &29.69 & 14.0 & 16.5 &
4& 4 & 0.6$\pm$0.3\\
17S& 07:00:20 & -11:23:20 & 9 & 1.31 & 4.3 &30.23 & 12.3 & 14.0 &
12& 5  & 0.6$\pm$0.3\\
37& 07:01:25 & -11:08:50 & 15 &2.1&5.3& 30.44 & 11.7 & 13.1 &
15& 5 & 0.4$\pm$0.6 \\
42& 07:01:34 & -11:11:16 & 10 &1.8 & 4.5 & 30.37 & 11.9 & 13.3 &
32& 9  & 0.4$\pm$0.6 \\
55S& 07:02:07 & -11:17:11 & 30 & 0.59 & 2.0 &29.88 & 13.4 & 15.6&
5& 3 & 1.2$\pm1.0$\\
55N & 07:02:08 & -11:16:15 & 30 & 0.72 & 2.6 & 29.97 & 13.1 & 15.2 &
2& 1 & 1.2$\pm1.0$\\
\hline
\end{tabular}
}
\label{tab1}
\end{center}
{\scriptsize
Columns description: (1) CMaX number; (2,3) coordinates: (4) diameter of the position error circle;
(5,6) count rate and  respective signal-to-noise ratio; (7) X-ray luminosity; (8, 9) Statistically expected near-IR 
and optical magnitudes calculated by adopting the correlation log$L_X$/absolute magnitude found for young stars 
(see Sect. 4.4); (10, 11) Number of possible optical counterparts within the X-ray contours, and number of corresponding 
objects found in the 2MASS catalogue; (12) adopted visual extinction based on a standard reddening law.

Notes: (a) The estimated errors are $\sigma_J$= 1.2 mag. and $\sigma_R$= 1.4 mag.;  (b)  
(less than 50\% of the candidates have near-IR counterpart).}
\end{table*}

The resulting X-ray maps, superimposed on the {\it Gemini} I-band images, are given in Figure 6 (as examples) and in Appendix B (for completeness). 
Two sources, CMaX-17 and CMaX-55, can be broken down into distinct sub-structures, 
for which the count rate was then integrated within the X-ray contours, using the same count-to-flux ratio as used for the original sources themselves 
(see above, Sect. 2). We name them CMaX-17S (South) and 17N (North), and similarly CMaX-55S and 55N. For three other sources, CMaX-15, CMaX-37, 
and CMaX-42,  no sub-structure is statistically significant.

Figure 6 illustrates two different cases, CMaX-15 and CMaX-17(N+S). Comparing  the X-ray contours and  the spatial distribution of the stars, 
we find that many (up to 30) relatively bright objects fall within the 
X-ray contours, indicating that one or more of these objects contributes to the X-ray emission. Conservatively, we first consider them all as candidate 
counterparts to the unresolved {\it ROSAT} sources.
Other objects are also visible on the optical image, but they are fainter and no reliable {\it Gemini} photometry could be obtained for them. 
Therefore, we restrict our analysis to objects brighter than R = 21 mag.

To further characterize candidate counterparts, we searched for near-IR sources within the {\it Gemini} fields using the 2MASS catalogue. These sources  
are shown  by crosses in Figure 6, while the relevant optical-near-IR identifications (i.e., within the X-ray contours) are indicated  by open squares. 
More than half of the visible
objects within the X-ray contours are too faint to appear in the 2MASS catalogue (limiting magnitude J = 17).
As a consequence, as shown in Table 1, only a fraction ( $\sim 46$\%) of the candidates have  VRIJHK data (6 of 12 for CMaX-15, 9 of 16 for CMaX-17, etc.).

We now have a first list of optical candidates selected on the basis of position (within 
the X-ray contours) and brightness (R $<$ 21).
But this list may still include foreground and background objects, and we have to find their intrinsic colours (using VRI photometry from {\it Gemini}, and JHK 
photometry from 2MASS when available), to establish their young-star nature, and to do so we must now consider interstellar extinction and IR excess 
(possibly indicative of circumstellar disks).

%========================================new Sect. 4.3

\subsection{Interstellar reddening}

The large-scale extinction models of Am\^ores \& L\'epine (2005), which are based on gas (HI and CO) and dust (IRAS) column densities, 
have a $\sim 0.5^o$ spatial resolution. They give $A_V$ = 0.4 mag for the foreground extinction in the direction of CMa R1. 
This is however a lower limit on a smaller spatial scale, since the  presence of CMa cloud must be also taken in account.
We have thus estimated the extinction towards the {\it ROSAT} sources  using star counts.

First we used the digitized POSS(R) image of the entire {\it ROSAT}  field with a resolution of $2.5'$ (Sect. 3).
More accurate estimates were kindly provided by Cambr\'esy (2002, private communication), who 
obtained an extinction map of the CMa R1 region based on source counts in the J band, from the near-IR {\it DENIS} Catalogue.
This method is the same as that used for the North America and Pelican nebulae (Cambr\'esy  et al. 2002).

%=======================New Fig. 7  J-H x H-K of all the Gemini candidates
\begin{figure}[!b]
\centering
\includegraphics[height=8cm,angle=0,clip]{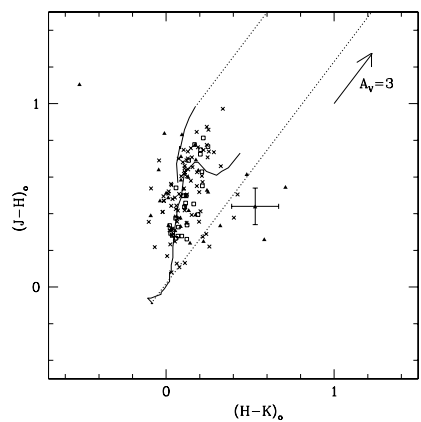}
\caption{Near-IR colour-colur diagram of the possible counterparts  of the X-ray sources 
studied in the {\it Gemini} fields (triangles). A typical error bar is shown for illustration. Two control groups of field stars 
are indicated by: (i) crosses for objects near ($\sim$ 1 arcmin) to the centre of the 
{\it Gemini} image, and (ii) open squares corresponding to stars located in a distant region free from obscuration. 
The full lines represent the Zero Age Main Sequence and Giant intrinsic colours. Interstellar 
reddening vectors are indicated by dashed line for B8V and M7III stars.
} 
\end{figure}
%%---------------------------------End Figure 7

Aiming to obtain the extinction in both optical (VRI) and near-IR  (JHK) bands, the conversion of $A_{\lambda}/A_V$ was done by 
adopting the reddening law from Cardelli et al. (1989) and the reddening-to-selective extinction ratio $R_V = 4$, which Terranegra et al. 
(1994) estimate in the direction of CMa R1. 
Last column of Table 1 gives the adopted visual extinctions and deviations of these values, measured in each {\it Gemini} field.

In order to check whether the magnitudes were correctly de-reddened,  field stars were selected in two control regions. A first region 
considers stars in a local area  ($1.5'$ radius around the  X-ray source centroid),
indicated by crosses in Figure 6 (excluding the candidates). The colours of the stars in this area 
were corrected by using the same extinction as adopted for the candidates.
The other region is located around $\alpha_{(J2000)} \sim $ 07$^h$, $\delta_{(J2000)} \sim$ -10$^o$
(upper right-hand corner of Fig. 1),
which is beyond the lowest IRAS contour and thus is free of cloud contamination.
In this case, the stars were de-reddened by using the foreground extinction  $A_J$=0.12 mag (equivalent to the value $A_V = 0.4$ 
mentioned previously from  Am\^ores \& L\'epine (2005)).
Figure 7 displays the  $(J-H)_o$ {\it vs.} $(H-K)_o$
diagram (de-reddened near-IR colours) for  the objects studied in the {\it Gemini} fields.
The intrinsic colours of the ZAMS stars (Siess et al. 2000) and giants  (Bessell et al. 1998) are also plotted.
The comparison with field stars of the control regions indicates that  the extinction corrections given in the last column of Table 1, can be
applied to the optical magnitudes of the {\it Gemini} candidates.

%%-------------------Figure 8 -------------------- Optical colours

\begin{figure}[!t]
\centering
\includegraphics[width=7cm,angle=0]{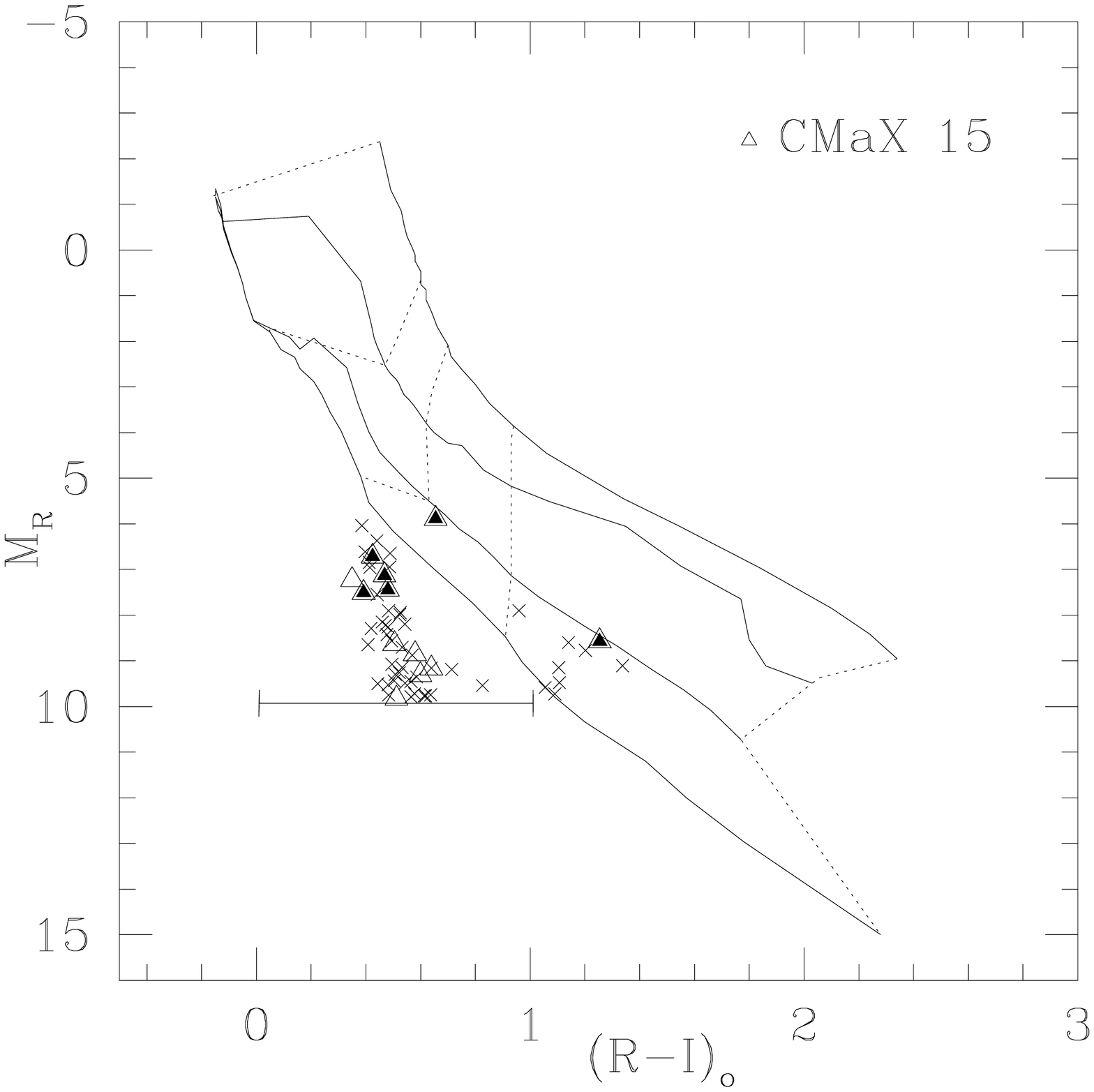}
\caption{Colour-magnitude diagram of the optical candidates to be counterparts of the 
X-ray source CMaX15. R- and I-band data 
were extracted from {\it Gemini} images and corrected 
for reddening using the extinction estimated for
each field (see text). An error bar typical of faint objects is shown for illustration.
 The field stars are indicated by crosses, while candidates are shown by triangles 
(filled symbols represent near-IR sources).
 The theoretical lines (intrinsic colours and isochrones) were
taken from Siess et al. (2000).  The
isochrones 0.1 Myr, 1Myr, and 10Myr, and the Zero Age Main Sequence (full lines) and evolutionary tracks for  0.1, 0.5, 1, 2,  
and 7 $M_{\odot}$ (dotted) are 
plotted. The absolute $M_R$ magnitude was obtained by adopting a distance 
modulus of 10 mag.}
\end{figure}
%%-------------------End of Figure 8 

%========================================Sect. 4.5

\subsection{Confirmed Candidates}

The intrinsic colours of the {\it Gemini} candidates were then analyzed for two different purposes :
(i) near-IR colour-colour diagrams, looking for H-K excesses; (ii) optical colour-magnitude diagrams, to confirm 
their young star nature and determine their masses and ages.

It can be noted in Fig. 7 and Fig. B2 (Appendix B) that three counterparts in {\it Gemini} fields show H-K excess. 
Even considering deviations on the corrections for extinction, the H-K excess is still high for these objects. This indicates 
the presence of circumstellar matter, as suggested for other four objects in Fields~1 and 2 (Sect. 3).
The relevant sources are indicated by an asterisk in Table A1.

The discussion of the optical characteristics of the counterparts in {\it Gemini} fields is based on the 
colour-magnitude diagram, using the R and I bands, and a distance modulus of 10 mag. Figure 8 
displays the absolute red magnitude ($M_{Ro}$)
as a function of $(R-I)_o$ (both corrected for extinction), compared to the theoretical models 
calculated for the ZAMS, 10 Myr, 1 Myr, and 0.1 Myr, 
in the mass range from 0.1 to 7 $M_{\odot}$ (Siess et al. 2000). 

Optical colours (Figure 8) and near-IR colours (Figure 7) were both used 
to select  the counterparts, based on colour excess. Therefore the objects having colours similar to the field 
stars (located in the left side of the ZAMS) 
were disregarded, while the objects  showing (H-K) or (R-I) excess (appearing above the ZAMS) were kept as counterparts. 
Adding this constraint, we can establish a final list of confirmed young star
counterparts, which is given in Table 2. Among the selected counterparts, CMaX42-1 is the only one that is not
associated to a near-IR source. In order to have a homogeneous sample of objects, we decided to exclude this
object of the overall analysis. This exclusion does not affect our results, since it is the faintest 
counterpart of CMaX42, giving a small contribution to the total X-ray luminosity, as explained below.  

The unresolved {\it ROSAT} sources in Field 1 are thus found to have up to two or four plausible counterparts each. There are two extreme possibilities: $(i)$ all the counterparts contribute more or less equally to the X-ray luminosity, 
or $(ii)$ only one dominates the X-ray  emission.

To do so, we compare the values of $L_X$ of the original {\it ROSAT} sources, with the ``integrated" J and R magnitudes of the confirmed 
counterparts, obtained by summing the individual J and R luminosities (derived from the individual J and R magnitudes), and converting 
back the result into magnitudes.
For a given CMaX source, we list in Table 2 the two extreme magnitudes (mentioned above): the ``integrated'' magnitudes R$_{int}$ and J$_{int}$, and
the individual magnitudes of the brightest candidate.

In Appendix  A we compare the integrated magnitudes to the X-ray luminosity. More precisely, we show in Figures A1 and A2 a diagram of  $L_X$ versus $M_{Ro}$
 and $M_{Jo}$. The {\it Gemini} ``integrated" candidates are found to lie on the high side, within $\sim 2 \sigma$ of the nominal correlation. This suggests that 
the X-rays are dominated by a small number of candidates (which would diminish the absolute magnitudes for a given X-ray luminosity). But since the evidence 
is not clear-cut, we will conservatively assume that the X-ray emission is spread among the candidates, in other words that all the candidates listed in Table 2 
(excepting CMaX42-1) are X-ray emitting young stars, to be added to the list of the resolved CMaX sources.

%%-----------------------------------------New Figure Fig.9
\begin{figure}[]
\includegraphics[height=8.5cm,angle=270]{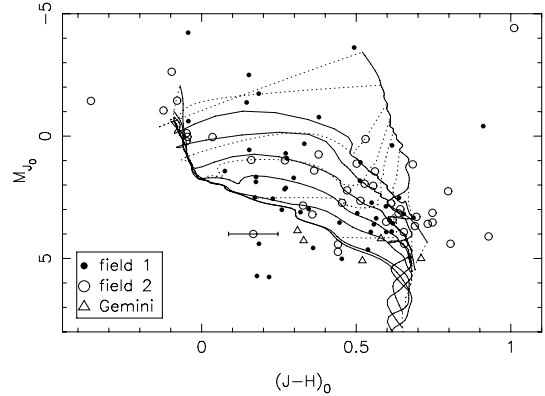}
\caption{Diagram of absolute J magnitude as a function of the J-H colour (both 
corrected for  extinction). The CMaX sources in Field~1 are represented by filled circles, sources in Field~2 are shown by
open circles. Five representative {\it Gemini} sources are shown by triangles.
The Zero Age Main Sequence (ZAMS)  and isochrones 
(0.1 to 20 Myr) are indicated by full lines and evolutive tracks (0.1 to 7 M$_{\odot}$) by dotted lines.}
\end{figure}
%%-----------------------------------------End Figure 9

%%%%%%%%%%%%%%%%%%%%%%%%%%%%%%%%%%End Sect.4%%%%%%%%%

%%========================================Sect. 5
\section{Stellar masses and ages}

The stellar masses and ages of the resolved CMaX sources (Table A1), and of the 
faint {\it Gemini} counterparts for the unresolved sources,
were derived from different colour-magnitudes diagrams, and compared with the pre-main sequence 
evolutionary models of Siess et al. (2000).
For the CMaX sources we used the near-IR M$_J$ vs. J-H diagram  (Fig.9). For the {\it Gemini} candidates we used the M$_R$ vs. R-I 
diagrams (Fig. 8 and Fig. B3), since magnitudes are too faint, making a comparison difficult with the  
isochrones in the near-IR colour-magnitude diagram.  Only near-IR isochrones are used in the comparison of age distribution of 
CMaX resolved sources in Fields ~1 and ~2.

From Figure 9, we find that 18 resolved CMaX sources are located ``outside" the isochrones (adopting error bars of $\pm 0.3$~mag in J-H), 
so we could determine masses and ages only for 40 of them in Field~1, and for 31 in Field~2. 
 The error bars of J-H color indicate that errors of 30\% to 50\%  can be expected for age and mass. For this reason, to estimate these parameters 
we decided to adopt ranges (or bins) defined by the isochrones separation (5 - 10 Myr, for example), instead of nominal values (7.5$\pm$2.5 Myr, for example).  
These determinations are listed in Table A1, except for sources marked with ``??''. 
Among the 18 objects with undetermined age and mass, 14 are located to the right side of the 0.1 Myr isochrone: this is probably due to the presence 
of circumstellar matter typical of YSOs.
On the other hand, four objects located to the left side of the 
isochrones are probably not the right counterparts of the CMaX sources.
Indeed, other faint objects are found within $30"$ of the center of 
the corresponding X-ray contours, but the lack of photometry for these 
four objects does not affect the age and mass distribution of CMaX 
sources.

Considering that counterparts identified in the {\it Gemini} fields have M$_J >$5 mag., most of them appear outside of the isochrones of the near-IR 
colour-magnitude diagram. Thus, Fig. 9 does not show the distribution of
all the {\it Gemini} counterparts, but, for illustration, only of the five brightest sources per field.  In this case, optical colours are more appropriate to 
determine the stellar parameters: their ages and masses  were estimated from M$_R$ vs. R-I diagrams shown in Figures 8 and B2 (Appendix B). 
From Fig. 9, one object (CMaX15-1) is approximately 20 Myr old and has a mass $\sim$ 1 M$_{\odot}$, 
%One object (CMaX15-1) has its age and mass also determined from Fig. 9,  giving approximately 20 Myrs and 1 M$_{\odot}$, 
in agreement with the result  (10-20 Myr, 0.5-1.0 M$_{\odot}$) obtained from Fig. 8 and indicated in Tab. 2.

We have compared our results with the cluster ages estimated by Soares \& Bica (2002, 2003; see above, Sect. 1 and 3). Soares \& Bica (2002) 
give an age of $\sim$ 1.5Myr for NGC2327 and BRC 27. Five resolved CMaX sources belong to these clusters, for which we have obtained 
similar ages. For the older cluster VdB-RN92 (5-7Myr, Soares \& Bica 2003) we  have identified four other CMaX sources, but only one of 
them (CMaX-68) has compatible age, while the others seem to be massive stars for which the age determination is uncertain. 

%%=============================New Table 2

%\begin{landscape}
\begin{table*}[ht] 
\caption{List of possible additional faint counterparts to CMaX sources.}
\smallskip
\begin{center}
{\scriptsize
\begin{tabular}{lcccccccccccc}
\hline
\noalign{\smallskip}
CMaX & R.A. & Dec. & R$_{int}$ & R & V-R & R-I & J$_{int}$ & J & J-H$^d$ & H-K$^d$ & Age & Mass \\
& J2000& J2000& (mag) & (mag) & (mag) &(mag)& (mag) & (mag) &(mag) & (mag)& (Myr) & (M$_{\odot}$) \\
\noalign{\smallskip}
\hline
\noalign{\smallskip}
15-1 & 7 00 13.364 & -11 18 53.43 & 16.5$^e$& 17.1$^a$ & 0.29 & 1.08$^b$ &14.4$^e$ &14.77 & 0.70 $\pm$ 0.06 & 0.19 $\pm$ 0.07 & 10-20 & 0.5-1.0 \\
15-2 & 7 00 13.509 & -11 19 11.14 & &18.3 & 0.70 & 0.78 & &16.84 & 0.66 $\pm$ 0.28 & 0.81 $\pm$ 0.28$^c$ & ?? & ?? \\
15-3 & 7 00 15.325 & -11 18 58.56 & &18.6 & 0.73 & 0.79 & &16.71 & 0.38 $\pm$ 0.29 & 0.68 $\pm$ 0.34$^c$ & ?? & ?? \\
15-4 & 7 00 15.378 & -11 19 3.78 & &19.7 & 1.6 & 1.56$^b$ & &16.45 & 0.76 $\pm$ 0.19 & 0.29 $\pm$ 0.24 &10-20 & 0.1-0.5 \\
\hline 
\noalign{\smallskip} 
17N & 7 00 18.724 & -11 22 24.02 & & 18.8 & 0.89 & 0.90$^b$ & &16.59 & 0.50$\pm$ 0.25 & 0.58$\pm$ 0.30$^c$  & ?? & ?? \\
17S1 & 7 00 18.806 & -11 23 16.10 & 16.9$^e$&17.1 & 0.9 & 0.79$^b$ &15.4$^e$ &15.58 & 0.77 $\pm$ 0.1 & 0.14 $\pm$ 0.14 & 20-50 & 0.5-1.0 \\
17S2 & 7 00 18.886 & -11 23 23.67 & &19.2 & 1.22 & 1.15$^b$ & &16.71 & 0.89 $\pm$ 0.22 & 0.14 $\pm$ 0.29 & 20-50 & 0.1-0.5 \\
\hline
\noalign{\smallskip}
37-1 & 7 01 22.297 & -11 09 9.29 & &17.7 & 0.9 & 0.97$^b$ & &16.18 & 0.65 $\pm$ 0.15 & -0.1 $\pm$ 0.24 & 10-20 & 0.5-1.0 \\
37-2 & 7 01 23.560 & -11 08 26.77 & &18.2 & 1.61 & 1.67$^b$ & &15.71 & 0.73 $\pm$ 0.12 & 0.07 $\pm$ 0.16 & 1-2 & 0.5-1.0 \\
37-3 & 7 01 24.356 & -11 09 03.00 & 15.9$^e$&16.3$^a$ & -0.3 & 0.88$^b$ &14.3$^e$ &14.78 & 0.55 $\pm$ 0.06 & 0.11 $\pm$ 0.08 & 5-10 & 0.5-1.0 \\
\hline 
\noalign{\smallskip} 
42-1 & 7 01 35.313 & -11 10 12.23 & &17.2 & 0.73 & 0.73$^b$ & && & & 20-50 & 0.5-1.0 \\
42-2 & 7 01 35.437 & -11 10 44.07 & 15.6$^e$&15.7$^a$ & -0.08 & 0.86$^b$ & 13.9$^e$ & 14.44 & 0.35 $\pm$ 0.06 & 0.08 $\pm$ 0.08 & 2-5 & 0.5-1.0 \\
42-3 & 7 01 35.473 & -11 10 14.46 & &16.4 & 0.74 & 0.71$^b$ & &14.70 & 0.73 $\pm$ 0.08 & 0.21 $\pm$ 0.08 & 10-20 & 0.5-1.0 \\
\hline
\noalign{\smallskip} 
55N & 7 02 8.021 & -11 16 17.65 & 18.2$^e$&18.2$^a$ & 0.98 & 1.08$^b$ &15.6$^e$ &15.67 & 0.64 $\pm$ 0.1 & 0.35 $\pm$ 0.13 & 10-20 & 0.1-0.5 \\
55S & 7 02 8.139 & -11 17 35.46 & &18.9  & 0.6 & 0.74 & &16.46 & 0.73 $\pm$ 0.2 & 0.58 $\pm$ 0.22$^c$ & ?? & ?? \\
\hline
\end{tabular}
}
\end{center}
Notes:(a) Magnitudes adopted from the NOMAD and/or USNO catalogues, when the estimation from
Gemini images are unavailable, due to saturation; (b) counterparts selected according to the R-I excess; (c) counterparts having H-K excess;
(d) The 2MASS Catalogue was used to obtain the JHK magnitudes, for which error bars are provided. 
(e) R$_{int}$ and J$_{int}$ represent the ``integrated'' magnitudes (see Sect. 4.2), which are shown
together with the brightest object of each group.
\end{table*}
%=======================End Table 2

%%----------------------Figure 10
\begin{figure}[!t]
\includegraphics[height=4.4cm,angle=270,clip]{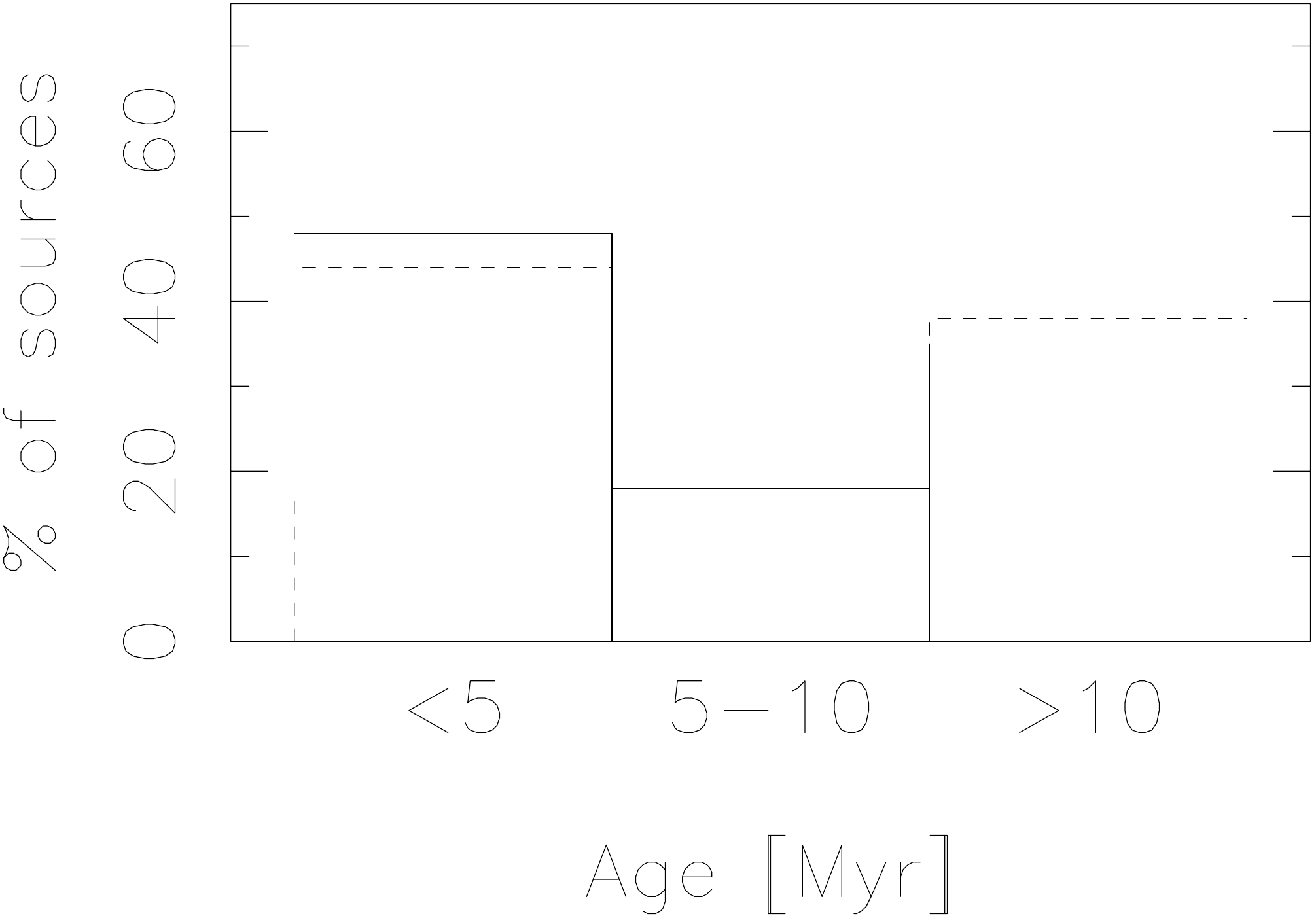}
\includegraphics[height=4.4cm,angle=270,clip]{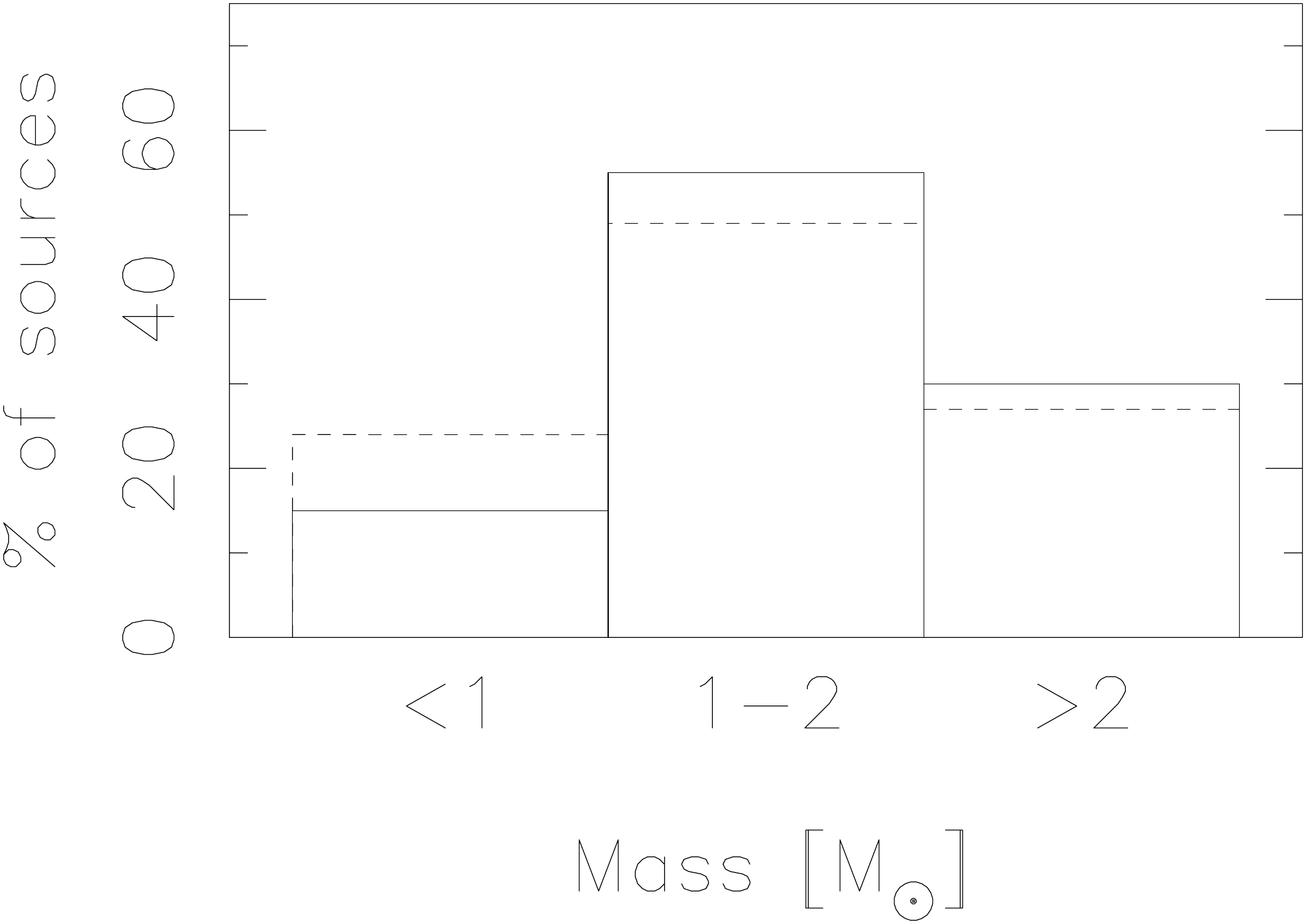}
\includegraphics[height=4.4cm,angle=270,clip]{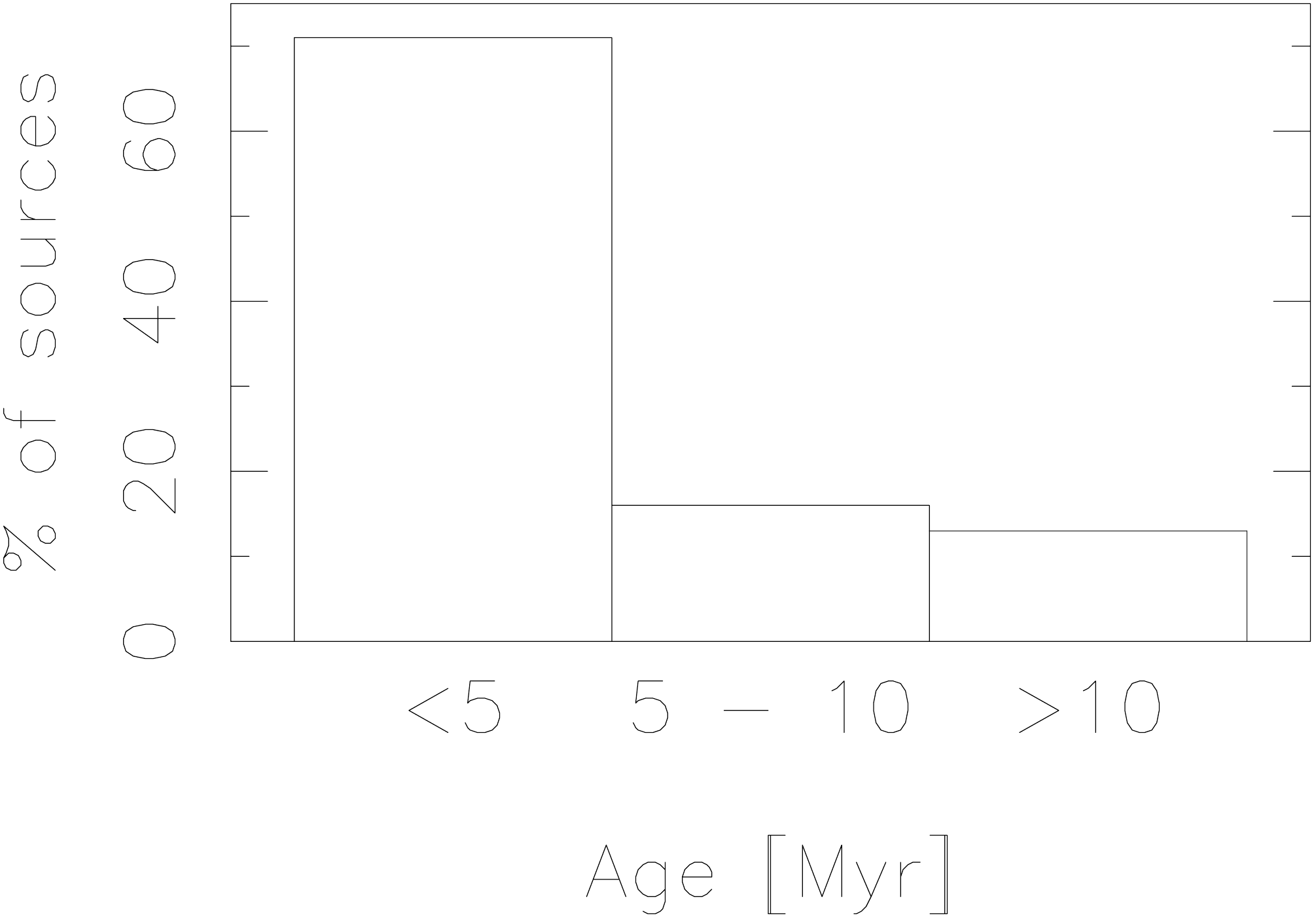}
\includegraphics[height=4.4cm,angle=270,clip]{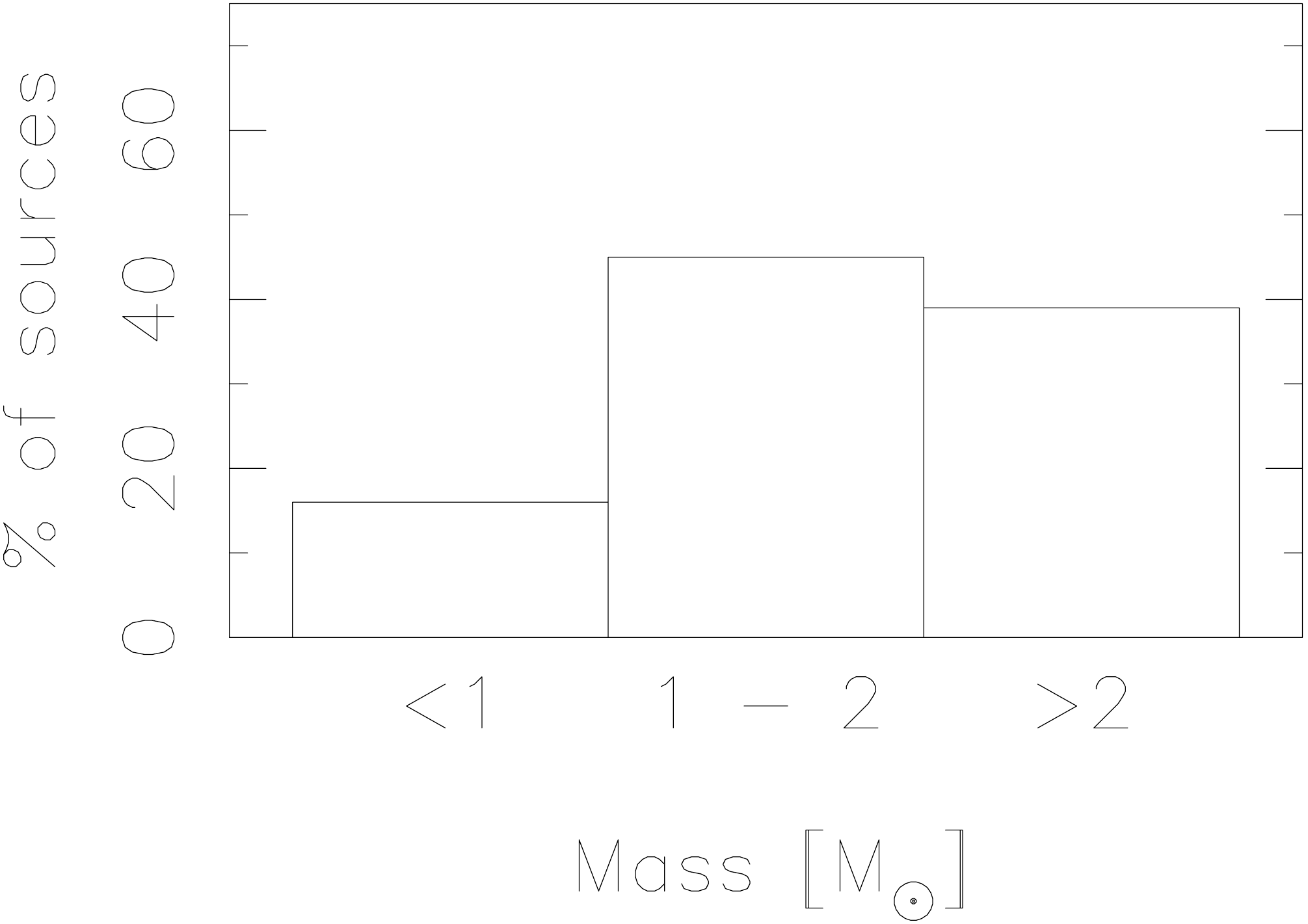}
\caption{Age and Mass distribution of the studied sample in Field 1 (top) and Field 2 (bottom).  Dashed lines are used to show 
the contribution of additional counterparts  in Field 1, detected by Gemini.}
\end{figure}
%%----------------------End Figure 10

Figure 10 presents the results, in the form of two panels, showing the broad distributions of ages and masses for all the CMaX sources in
Field~1 and Field~2 respectively. 
The percentage of sources for the each bin was obtained by dividing
the number of objects in the bin by the total number of sources in each field.

As concerns the {\it Gemini} candidates, two possibilities are explored in the histograms (for Field~1; see Table 2), considering: (i) only the 40 counterparts 
of resolved CMaX sources (indicated by full lines in Fig. 10);    (ii)  the brightest {\it Gemini} stars as counterparts to unresolved CMaX sources, 
leading to 40 + 5 objects (dashed lines in Fig. 10).
Note that objects outside the isochrones of Figure 9 have not been included in the distributions of ages and masses.

We comment on these results in the next section.

%%----------------------Figure 11
\begin{figure*}[ht]
\includegraphics[height=8cm,angle=270]{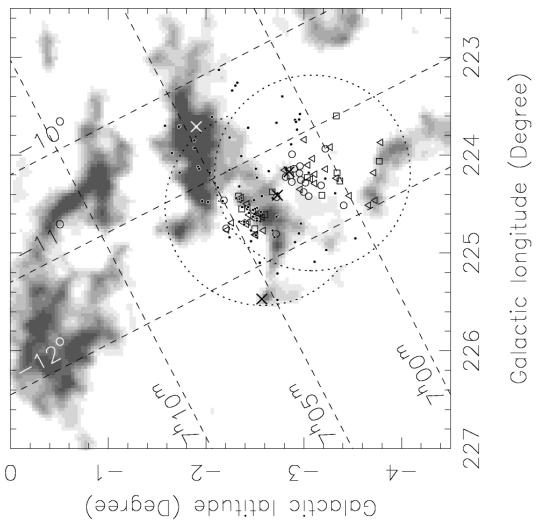}
\includegraphics[height=8cm,angle=270]{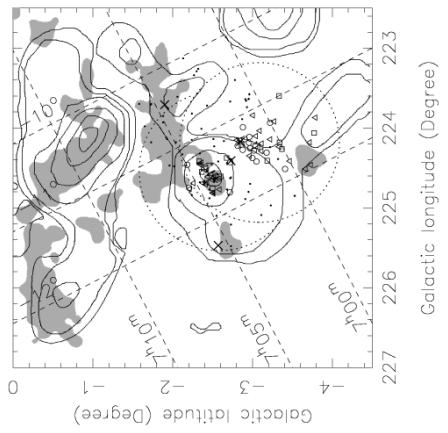}
\caption{Spatial distribution of the CMaX sources, in galactic coordinates, compared to: (left)  visual extinction (Dobashi et al. 2005) and (right)  
CO maps: $^{13}$CO is indicated by grey areas and $^{12}$CO by full line contours. 
The 2$^o$ circles (dotted lines) 
show the {\it ROSAT} fields. The different age bins of Figure 10 are represented by triangles ($<$ 5Myr), squares (5-10 Myr), and circles 
($>$ 10 Myr). The position of emission line stars detected by Shevchenko et al. (2004) is shown by dots. Large crosses indicate 
Sh-292, GU~CMa, FZ~CMa, Sh-297 (ordered by increasing galactic longitude).}
\end{figure*}
%%---------------------- End Figure 11

%%========================================New Sect. 6

\section{Conclusions}

\subsection{Summary of results}

One way to elucidate the nature of the ionized ``CMa R1 ring'' or ``Arc'' (Sh2-296) is to investigate the stellar population in its vicinity, more precisely 
to reconstruct the star formation history of the region. Until the present work, only the bright-star end of the population (from early B stars to F and K 
giants), i.e., the high-mass/young population, was surveyed. 
In the present paper, we used the most powerful tool to find the low-mass, older late-type population,  namely their X-ray emission resulting from 
magnetic activity. To this end, we analyzed or reanalyzed archival {\it ROSAT} data. 

Two overlapping {\it PSPC} fields exist, which cover a wide area, over nearly 5 sq. deg., i.e., the equivalent of a mosaic of 20 {\it XMM-Newton} fields, or 80 {\it Chandra} fields. The price to pay, however, compared to these satellites, is a relatively modest sensitivity, especially in one of the two fields which has a short exposure. 
More precisely, we find that our {\it ROSAT} survey is limited to $M_\star \sim 0.5 M_\odot$, as opposed to the usual limit of $M_\star \sim 0.1 M_\odot$ 
easily reached for typical {\it XMM-Newton} or {\it Chandra} observations of star-forming regions.

The two {\it ROSAT} fields (see Fig.1 and Fig.2), labelled by order of increasing Right Ascension, are ``Field 1'', west of the CMa ring (20 ksec exposure), 
and ``Field 2'', centred on the Herbig star Z CMa and overlapping most of the ring but with a short exposure (5 ksec). Fifty-six sources are detected in 
Field~1, and 37 in Field~2, 5 more being detected in the overlap area between Field~1 and Field 2. Note that, since the exposure is 4 times shorter in
 Field~2 than in Field~1, the expected number of sources, if similar in nature, would be 2 times less, i.e., 28: therefore, X-ray sources seem more 
numerous in Field 2 relative to Field 1, although the difference is marginally within Poissonian errors. In all, $\sim$ 100 sources are detected.

To identify the {\it ROSAT} sources, we searched catalogues (USNO and 2MASS), and digitized plates (POSS(R)), in order to characterize them via 
their colours. Not too surprisingly, most of the X-ray sources have no known counterpart, and we used our own photometry from the POSS(R) plate (Field 1), 
or USNO magnitudes when unavailable (Field 2), supplemented by 2MASS data. For five sources having no counterpart   inside the position error circle, we 
observed the corresponding fields with {\it Gemini} South to find possible fainter counterparts and determine their VRI magnitudes. The spatial 
distribution of the sources suggests the existence of two distinct X-ray clusters. One is visible in Field 2, closely associated with Z CMa and the 
CMa ring, that includes previously known optical clusters studied by Soares \& Bica (2002, 2003), and which we will name globally the ``Z CMa 
cluster'' hereafter; and a new one, discovered in Field 1, west of the CMa ring, which includes the bright star GU CMa (B2V), spatially distinct 
from the previous one.
We name this newly discovered cluster the ``GU CMa cluster''.

The presence of circumstellar matter is suggested for 7 counterparts that show H-K excess. This excess does not affect the estimation of stellar 
parameters such as mass and age, which were derived from M$_{J_o}$ {\it vs.} (J-H)$_o$ diagram.
Three figures summarize our X-ray source characterization. Figure 9 shows a colour-magnitude diagram for our sample, and PMS evolutionary 
tracks (Siess et al. 2000). This yields mass and age estimates for the sources, which are summarized in broad bins in Figure 10 for the two 
clusters separately: masses $<1, 1-2, >2 M_\odot$, ages $<5, 5-10, >10$ Myr. Figure 11 locates the position of the {\it ROSAT} sources with 
respect to the molecular gas of the region, traced by $^{12}$CO, and $^{13}$CO for the densest regions, compared to the visual extinction map 
from Dobashi et al. (2005).

\subsection{Discussion}

Although the error bars on masses and ages are large  ($\sim$ 30\%), Fig.10 shows significant similarities and differences 
between the Z CMa and GU CMa clusters.

$(i)$ The mass functions above $1 M_\odot$ are similar, with a maximum in the $1-2 M_\odot$ range. Below $1 M_\odot$, however, our sample 
is incomplete, essentially because it misses the X-ray faint, low-mass stars. This is exemplified by the {\it Gemini} sources in the GU CMa cluster: 
they are mostly detected as groups, and thus tend to fill the $< 1 M_\odot$ mass bin. For both clusters, this mass bin should therefore be considered 
only as indicative of the low-mass, faint population.

$(ii)$ The age distribution is more significant. The GU CMa cluster appears older than the Z CMa cluster: 35-38\% of the GU CMa members (depending 
on whether or not the {\it Gemini} sources are taken into account) are older than 10 Myr, whereas nearly 70\% of the Z CMa cluster stars 
are younger than 5 Myr. (Conversely, $\sim 45$\% of the GU CMa members are younger than 5 Myr, and 13\% of the Z CMa members are older than 
10 Myr: we will return to this point below). In both clusters, the intermediate-age population is similar ($\sim 17$\% between 5 and 10 Myr).

$(iii)$ The most important difference is the space distribution of the X-ray sources, compared with the gas ($^{13}$CO: Kim et al. 2004; $^{12}$CO, 
May et al. 1988) and dust (Dobashi et al. 2005) distributions (Figure 11; see also Figures 1 and 2). The Z CMa cluster X-ray sources are extremely well correlated 
with the CMa ring and $^{13}$CO emission from dense gas, consistent with the high fraction of these sources being very young. 
This demonstrates that an active star formation episode is currently going on, likely induced by the ring, especially if the X-ray source number excess in Field 2 over Field 1 
noted above is real. This qualitatively confirms previous optical and IR studies, but adds several tens  
of young low-mass stars to the existing census.
In contrast, the GU CMa cluster is located in a ``vacuum'', comparatively far from the dense $^{13}$CO gas, just at the limit of the lower-density $^{12}$CO gas. 
This is also consistent with the older age of this cluster. 

We thus have now evidence for two distinct episodes of star formation,  as previously suspected by Schevchenko et al. (1999) for a couple of  
bright stars (GU CMa and FZ CMa): star formation in the Z CMa region is  currently going on, fueled by dense molecular material at the edge of 
the CMa ring, but, in the absence of such material, must have ceased several Myr ago in the GU CMa cluster.

Yet this evidence is not clearcut: why is there a significant fraction (16\%) of old stars in the Z CMa region (a previously known result, as recalled 
in the Introduction, which we confirm and extend on the basis of lower-mass stars)? And conversely, why is there a significant fraction of young 
stars (44\%) in the GU CMa cluster? A possible answer is a mixing, due to stellar proper motions, of the two clusters: at a distance of $\sim 1$ kpc, 
the centroids of the two clusters would be $\sim 17$ pc apart ($\sim 1^o$ on the sky). Mixing could start to be visible if the stars had a proper motion 
of order a few km~s$^{-1}$, which is reasonable, provided the centroids themselves would not have moved significantly. The main objection against 
such an interpretation is the lack of a ``well-mixed'' population between the two clusters, which appear spatially very distinct. However, there may be 
some extinction effect: the ``intercluster'' region is within a low-density, but extended region visible in $^{12}$CO (Figure 11) and in the IRAS data 
(Figure 1), possibly yielding column densities high enough to hide faint X-ray sources. More sensitive X-ray observations would be needed to 
clarify this point.

A related question is that of the connection between the newly discovered GU CMa cluster, and the surrounding molecular clouds. Figure 11 shows 
that the GU CMa cluster lies in a cavity, roughly symmetrical (with respect to $^{13}$CO contours) to the cavity associated with the CMa ring. This 
could suggest that the GU CMa cluster cavity has been excavated by stellar winds and/or supernovae from the now defunct high-mass end of the 
cluster stars. No diffuse X-ray emission is visible in our {\it ROSAT} PSPC data, but this is not too surprising in view of the sensitivity required to 
detect this emission (see the example of Orion with {\it XMM-Newton}, G\"udel et al. 2008). Some diffuse X-ray emission does appear to exist in 
this region in the 3/4 keV {\it ROSAT} diffuse background (Snowden et al. 1997), but it is on a large scale and not directly related to the CMaX 
region in particular. 
An interesting test of the stellar wind excavation hypothesis would be  to obtain a census of circumstellar disks (in the near-IR range)  
around the youngest stars of the cluster (i.e., with ages $< 5$ Myr and noted by  triangles in Figure 11), which, under normal conditions should have  
still retained their original disk (e.g., Dahm \& Hillenbrand 2007).  
Indeed, stellar winds in OB associations are known to be able to  quickly blow away disks, as is the case, for instance, in the Sco-Cen
association (Preibisch \& Zinnecker 1999). Along these lines, we note that this hypothesis is indeed consistent with the very small number of sources
 with near-IR excess in the GU CMa cluster.

\subsection{Concluding remarks}

By analyzing archival {\it ROSAT} data, which give access to the previously unknown low-mass ($M_\star \sim 0.5 M_\odot$), old stellar population, 
we have shown in this paper that star formation in the CMa R1 region has been going on for more than 10 Myr, but in two separate episodes, each giving rise to a 
distinct cluster. The older one, which we name the ``GU CMa" cluster, has been discovered only by way of X-ray observations, supplemented by 
our photometric characterization of the X-ray sources. Given its location far from dense molecular material, it is clear  that star formation has now 
ceased in this cluster. The existence of a younger cluster, which is observed around the Herbig star Z CMa, has been 
known for some time as a result of previous surveys, but our X-ray observations add several tens of low-mass stars to the existing census. A 
puzzling question remains, which is the presence in both clusters of a small, but significant fraction of young ($< 5$ Myr) and older ($> 10$ Myr) 
stars. Spatial mixing between the two clusters is a possibility, but more sensitive X-ray observations would be needed (to lower the mass limit and/or 
overcome the extinction) to clarify this point.

At any rate, there does not seem to be any connection between the GU CMa cluster and the CMa ring. The star formation episode that gave rise to 
GU CMa and its cluster likely predates the ring, perhaps by as much as a few Myr. Only the clusters in the Z CMa region appear related to the ring. 
It is entirely possible that other older clusters like the GU CMa cluster exist in the region, in particular inside the ring, east of Z CMa. As demonstrated 
with the discovery of the GU CMa cluster, observations with wide-field X-ray telescopes like {\it ROSAT} are very powerful to detect such clusters, 
but unfortunately that era is over, and even with a moderate field-of-view like the EPIC camera aboard {\it XMM-Newton}, a mosaic of many fields 
is required to map large areas. Until this is done, the origin of the CMa ring will likely remain an unsolved problem.

%%==================================================

\begin{acknowledgements}
We would like to express our gratitude to A. Hetem Jr. for his kindness preparing some of the figures (composite maps) that improved the 
presentation of the paper.
We are grateful to J. Guibert for help in acquiring the data
from the plate digitizing machine MAMA.
The authors thank FAPESP for partial financial support (JGH: Procs. No. 2001/09018-2
and No. 2005/00397-1; CVR: Proc 01/12589-1). JGH and TM thank support from
USP/COFECUB (Proc. No. 2007.1.435.14.5) and Fapesp/CNRS (Proc. No. 2006/50367-4).
This work has made use of the SIMBAD, VizieR, and
Aladin databases operated at CDS, Strasbourg, France.
This publication makes use of data products from the Two Micron All Sky Survey, which is a joint project of the 
University of Massachusetts and the Infrared Processing and Analysis Center/California Institute of Technology, 
funded by the National Aeronautics and Space Administration and the National Science Foundation.

\end{acknowledgements}

\begin{appendix} %First online appendix --------------------------A

\section{X-ray emission in CMaR1}

%%------------------------------------------------------------Section A.1
\subsection{Sources detected in CMa R1}
The {\it ROSAT}  PSPC $2^o$ image of Field~1 was obtained with an exposure of 20~ksec and 
pointed towards 
$\alpha$ = 07$^h$ 01$^m$, $\delta$ = -11$^o$ 24' (J2000 coordinates).
 Due to a strong background emission of instrumental and cosmic 
origin in the spectral range 0.1-0.4 keV (see Gregorio-Hetem et al. 1998), the 
sources were analyzed in the 0.4-2.4 keV range. The  X-ray data derived from the {\it ROSAT} image analysis 
is given in Table A1, along with stellar identifications.

Sixty-one X-ray sources were detected, 48 of them having S/N $>$ 3.5, with X-ray luminosities in the range of  $6 \times 10^{30}$ erg~s$^{-1}$.
to $8 \times 10^{32}$ erg~s$^{-1}$. Following Gregorio-Hetem et al.  (1998) the approximate X-ray luminosities (\~L$_X$) were derived by  
using the correspondence between count-rate and X-ray flux given by 1 cnt~ks$^{-1}$ $\simeq 9 \times 10^{-15}$ erg~s$^{-1}$~cm$^{-2}$. The  
assumed visual extinction, temperature and distance are respectively $A_V$=1 mag, $kT_X$=1 keV, and $d \sim$ 1 kpc for all the sources.

A similar procedure was adopted to obtain \~L$_X$ for the sources detected in the
{\it ROSAT} field observed by Zinnecker \& Preibisch (1994). This other PSPC image
was pointed to  $\alpha$ = 07$^h$ 03.7$^m$, $\delta$ = -11$^o$ 33' and had an
exposure of 5~ksec. We have identified 43 sources in this field, five of them coinciding
with the sources identified in the first field (described above). The  range of X-ray luminosities
in this sample  is $10^{30}$ erg~s$^{-1}$ to $9.8 \times 10^{30}$ erg~s$^{-1}$.

%%-----------------------------Table A1
\begin{table*}[!t]
\caption{X-ray sources detected in CMa R1 by {\it ROSAT}}
\smallskip
\begin{center}
{\scriptsize
\begin{tabular}{cccccccccccccc}
\hline
\noalign{\smallskip}
CMaX &	Err&Cnt/ks	&			S/N&		logLx &	R.A.  	&	Dec.  	&	R	&	$A_V$	&	J	  &	H	     &	K	       &	Age	&	Mass	\\
         &    (")      &         &               &             & J2000  &  J2000    & (mag) &  (mag)  &   (mag) &   (mag) &  (mag)   &  (Myr)    &   ($M_{\odot}$)  \\
\noalign{\smallskip}
\hline
\noalign{\smallskip}
1	&	30	&	16.7	$\pm$	1.6	&	10.4	&	31.25	&	6 57 52.6	&	-11 27 23	&	9.7	&	1.0	&	8.9	&	8.7	&	8.6	&	$<$1	&	3.0-5.0	 \\
2	&	30	&	5.0	$\pm$	0.8	&	6.3	&	30.72	&	6 58 17.6	&	-11 37 10	&	15.7	&	1.0	&	14.5	&	13.9	&	13.8	&	05-10	&	1.0-1.5	 \\
3$^a$	&	30	&	17.9	$\pm$	1.6	&	11.2	&	31.28	&	6 58 42.6	&	-11 42 00	&	9.0	&	0.5	&	8.4	&	8.2	&	8.1	&	$<$1	&	3.0-5.0	 \\
4$^b$	&	30	&	19.4	$\pm$	1.4	&	13.9	&	31.31	&	6 59 02.0	&	-11 00 38	&	10.8	&	0.4	&	11.5	&	11.4	&	11.4	&	05-10	&	1.5-2.0	 \\
5$^c$	&	30	&	5.0	$\pm$	0.9	&	5.6	&	30.72	&	6 59 10.7	&	-11 40 03	&	8.3	&	0.5	&	7.7	&	7.5	&	7.4	&	$<$1? 	&	$>$5	 \\
6	&	30	&	7.0	$\pm$	0.9	&	7.8	&	30.87	&	6 59 10.7	&	-11 57 30	&	13.0	&	1.0	&	11.2	&	10.8	&	10.7	&	02-05	&	2.0-3.0	 \\
7	&	30	&	5.1	$\pm$	0.8	&	6.4	&	30.73	&	6 59 31.0	&	-11 58 18	&	11.8	&	0.9	&	11.0	&	10.6	&	10.6	&	02-05	&	2.0-3.0	 \\
8	&	20	&	2.4	$\pm$	0.6	&	4.0	&	30.40	&	6 59 51.1	&	-11 15 52	&	12.7	&	0.5	&	11.2	&	10.7	&	10.5	&	01-02	&	2.0-3.0	 \\
9	&	12	&	1.4	$\pm$	0.4	&	3.5	&	30.17	&	(7 00 00.5)	&	(-11 14 09)	&	--	& 	--	&	--	&	--	&	--	&	--	&	--	 \\
10	&	30	&	5.1	$\pm$	0.7	&	7.3	&	30.73	&	7 00 03.4	&	-11 47 45	&	--	&	--	&	9.6	&	9.6	&	9.6	&	$<$ 1	&	3.0-5.0	 \\
11	&	10	&	2.8	$\pm$	0.5	&	5.6	&	30.47	&	7 00 03.8	&	-11 15 26	&	14.8	&	0.4	&	13.7	&	13.2	&	13.0	&	20-50	&	1.0-1.5	 \\
12	&	5	&	3.7	$\pm$	0.6	&	6.2	&	30.59	&	7 00 05.2	&	-11 31 46	&	14.8	&	0.8	&	13.4	&	12.8	&	12.7	&	05-10	&	1.0-1.5	 \\
13	&	12	&	1.2	$\pm$	0.3	&	4.0	&	30.10	&	7 00 08.0	&	-11 37 00	&	14.9	&	0.5	&	14.1	&	13.4	&	13.2	&	05-10	&	1.0-1.5	 \\
14	&	12	&	2.2	$\pm$	0.4	&	5.5	&	30.36	&	7 00 12.5	&	-11 34 33	&	16.6	&	0.8	&	13.8	&	13.1	&	12.9	&	02-05	&	1.0-1.5	 \\
15	&	10	&	1.9	$\pm$	0.4	&	4.8	&	30.30	&	(7 00 14.5)	&	(-11 18 51)	&	--	&	--	&	--	&	--	&	--	&	--	&	--	 \\
16	&	10	&	2.0	$\pm$	0.5	&	4.0	&	30.32	&	7 00 17.0	&	-11 08 03	&	13.7	&	0.4	&	12.6	&	12.4	&	12.3	&	$>$50 	&	1.5-2.0	 \\
17	&	9	&	1.3	$\pm$	0.3	&	4.3	&	30.14	&	(7 00 19.8)	&	(-11 23 20)	&	--	&	--	&	 - -	&	--	&	--	&	--	&	--	 \\
18	&	8	&	1.1	$\pm$	0.3	&	3.7	&	30.06	&	7 00 26.5	&	-11 26 33	&	14.9	&	0.8	&	13.1	&	12.4	&	12.2	&	02-05	&	1.0-1.5	 \\
19	&	30	&	4.6	$\pm$	0.8	&	5.8	&	30.69	&	7 00 27.1	&	-11 51 26	&	13.4	&	0.8	&	12.4	&	12.1	&	12.0	&	10-20	&	1.5-2.0	 \\
20$^d$	&	5	&	62.7	$\pm$	1.9	&	33.0	&	31.82	&	7 00 37.6	&	-11 14 45	&	7.8	&	0.7	&	6.6	&	6.0	&	5.9	&	$<$1?	&	$>$5	 \\
21	&	30	&	41.6	$\pm$	1.9	&	21.9	&	31.64	&	7 00 40.4	&	-11 04 35	&	12.3	&	0.9	&	10.6	&	9.9	&	9.7	&	$<$1	&	1.0-1.5	 \\
22	&	6	&	2.0	$\pm$	0.4	&	5.0	&	30.32	&	(7 00 41.3)	&	(-11 34 50)	&	--	&	--	&	--	&	--	&	--	&	--	&	--	 \\
23	&	5	&	3.4	$\pm$	0.5	&	6.8	&	30.55	&	7 00 43.4	&	-11 17 14	&	14.3	&	0.5	&	12.9	&	12.3	&	12.2	&	02-05	&	1.5-2.0	 \\
24$^*$	&	8	&	1.1	$\pm$	0.4	&	2.8	&	30.06	&	7 00 45.4	&	-11 31 00	&	16.9	&	0.5	&	15.9	&	15.6	&	15.0	&	??	&	??	 \\
25	&	8	&	0.4	$\pm$	0.2	&	2.0	&	29.62	&	7 00 45.5	&	-11 20 29	&	17.0	&	0.8	&	16.0	&	15.7	&	15.6	&	??	&	??	 \\
26	&	12	&	1.2	$\pm$	0.3	&	4.0	&	30.10	&	7 00 54.6	&	-11 34 12	&	14.0	&	0.8	&	13.2	&	12.8	&	12.7	&	20-50	&	1.0-1.5	 \\
27$^*$	&	10	&	2.2	$\pm$	0.5	&	4.4	&	30.36	&	7 00 55.5	&	-11 21 45	&	15.7	&	0.7	&	14.6	&	14.3	&	14.1	&	01-02	&	0.1-0.5	 \\
28	&	8	&	0.9	$\pm$	0.3	&	3.0	&	29.98	&	7 00 55.8	&	-11 26 25	&	15.8	&	0.8	&	13.6	&	12.8	&	12.6	&	??	&	??	 \\
29	&	5	&	2.5	$\pm$	0.4	&	6.3	&	30.42	&	7 01 00.8	&	-11 19 29	&	15.4	&	1.0	&	12.8	&	12.1	&	11.9	&	$<$1	&	0.5-1.0	 \\
30	&	12	&	2.1	$\pm$	0.5	&	4.2	&	30.34	&	7 01 04.1	&	-11 39 48	&	16.4	&	0.7	&	15.2	&	14.4	&	14.2	&	05-10	&	0.1-0.5	 \\
31	&	9	&	0.8	$\pm$	0.3	&	2.7	&	29.93	&	7 01 06.9	&	-11 31 23	&	10.6	&	0.7	&	9.4	&	9.0	&	8.8	&	$<$1	&	3.0-5.0	 \\
32	&	8	&	2.1	$\pm$	0.4	&	5.3	&	30.34	&	7 01 11.2	&	-11 27 50	&	11.4	&	0.9	&	10.8	&	10.6	&	10.5	&	02-05	&	2.0-3.0	 \\
33	&	15	&	0.9	$\pm$	0.3	&	3.0	&	29.98	&	7 01 13.1	&	-11 21 01	&	16.4	&	1.0	&	14.9	&	14.4	&	14.3	&	??	&	??	 \\
34	&	8	&	1.3	$\pm$	0.3	&	4.3	&	30.14	&	7 01 16.3	&	-11 25 58	&	15.2	&	1.0	&	13.7	&	13.0	&	12.9	&	05-10	&	1.0-1.5	 \\
35	&	8	&	0.7	$\pm$	0.3	&	2.3	&	29.87	&	7 01 20.1	&	-11 17 44	&	15.6	&	0.7	&	13.3	&	12.9	&	12.8	&	20-50	&	1.0-1.5	 \\
36	&	15	&	1.1	$\pm$	0.3	&	3.7	&	30.06	&	7 01 24.4	&	-11 09 06	&	16.2	&	0.5	&	14.8	&	14.2	&	14.1	&	20-50	&	0.5-1.0	 \\
37	&	15	&	2.1	$\pm$	0.4	&	5.3	&	30.34	&	(7 01 25.0)	&	(-11 08 50)	&	--	&	--	&	--	&	--	&	--	&	--	&	--	 \\
38	&	5	&	3.5	$\pm$	0.4	&	8.8	&	30.57	&	7 01 26.8	&	-11 28 21	&	13.5	&	0.8	&	12.3	&	12.0	&	11.9	&	10-20	&	1.5-2.0	 \\
39	&	8	&	0.8	$\pm$	0.3	&	2.7	&	29.93	&	7 01 29.2	&	-11 21 30	&	15.4	&	0.8	&	14.1	&	13.5	&	13.5	&	10-20	&	1.0-1.5	 \\
40	&	10	&	2.0	$\pm$	0.4	&	5.0	&	30.32	&	7 01 30.4	&	-11 22 29	&	16.9	&	1.2	&	15.4	&	14.8	&	14.8	&	??	&	??	 \\
41	&	8	&	1.6	$\pm$	0.4	&	4.0	&	30.23	&	7 01 34.0	&	-11 25 33	&	15.3	&	0.8	&	13.8	&	13.2	&	13.1	&	10-20	&	1.0-1.5	 \\
42	&	10	&	1.8	$\pm$	0.4	&	4.5	&	30.28	&	(7 01 34.3)	&	(-11 11 16)	&	--	&	--	&	--	&		--&	--	&	--	&	--	 \\
43	&	15	&	0.6	$\pm$	0.3	&	2.0	&	29.80	&	07 01 35.1	&	-11 36 38	&	14.3	&	1.3	&	12.1	&	11.8	&	11.7	&	10-20	&	1.5-2.0	 \\
44	&	8	&	1.2	$\pm$	0.3	&	4.0	&	30.10	&	7 01 35.8	&	-11 17 36	&	13.6	&	0.8	&	12.8	&	12.5	&	12.4	&	10-20	&	1.0-1.5	 \\
45	&	10	&	1.2	$\pm$	0.4	&	3.0	&	30.10	&	7 01 41.1	&	-11 32 57	&	12.9	&	1.0	&	12.2	&	11.9	&	11.9	&	10-20	&	1.5-2.0	 \\
46	&	15	&	1.4	$\pm$	0.4	&	3.5	&	30.17	&	(7 01 43.0)	&	(-11 10 20)	&	--	&	--	&	--	&	--	&	--	&	--	&	--	 \\
47	&	8	&	2.1	$\pm$	0.4	&	5.3	&	30.34	&	7 01 45.9	&	-11 30 33	&	15.4	&	0.7	&	13.4	&	12.7	&	12.5	&	01-02	&	0.5-1.0	 \\
48$^e$	&	15	&	2.7	$\pm$	0.4	&	6.8	&	30.45	&	7 01 49.3	&	-11 18 07	&	7.5	&	1.7	&	6.3	&	6.2	&	6.0	&	??	&	??	 \\
49	&	9	&	1.5	$\pm$	0.4	&	3.8	&	30.20	&	7 01 49.5	&	-11 16 41	&	12.2	&	1.7	&	10.8	&	10.3	&	10.2	&	02-05	&	2.0-3.0	 \\
50	&	12	&	1.1	$\pm$	0.3	&	3.7	&	30.06	&	7 01 52.4	&	-11 20 18	&	13.4	&	1.3	&	12.1	&	11.7	&	11.5	&	05-10	&	1.5-2.0	 \\
51	&	8	&	1.3	$\pm$	0.4	&	3.3	&	30.14	&	7 01 56.3	&	-11 24 07	&	16.5	&	1.2	&	14.3	&	13.6	&	13.4	&	10-20	&	1.0-1.5	 \\
52	&	8	&	2.3	$\pm$	0.4	&	5.8	&	30.38	&	7 01 58.7	&	-11 15 45	&	13.5	&	1.7	&	12.3	&	11.6	&	11.5	&	01-02	&	2.0-3.0	 \\
53	&	14	&	1.5	$\pm$	0.4	&	3.8	&	30.20	&	7 02 04.4	&	-11 19 44	&	14.1	&	1.2	&	13.4	&	13.0	&	12.9	&	$>$50	&	05-1.0	 \\
54	&	15	&	0.5	$\pm$	0.3	&	1.7	&	29.72	&	(7 02 05.5)	&	(-11 26 50)	&	--	&	--	&	--	&	--	&	--	&	--	&	--	 \\
55	&	30	&	0.6	$\pm$	0.3	&	2.0	&	29.80	&	(7 02 07.4)	&	(-11 17 11)	&	--	&	--	&	--	&	--	&	--	&	--	&	--	 \\
56	&	30	&	5.3	$\pm$	0.8	&	6.6	&	30.75	&	7 02 38.9	&	-11 30 39	&	12.8	&	0.9	&	9.9	&	8.9	&	8.6	&	??	&	??	 \\
\hline																													
57$^f$	&	30	&	2.3	$\pm$	1.0	&	2.3	&	30.40	&	7 02 42.6	&	-11 27 12	&	8.0	&	1.6	&	7.8	&	7.8	&	7.7	&	$<$1?	&	$>$5	 \\
58	&	30	&	2.0	$\pm$	1.0	&	2.0	&	30.32	&	7 02 47.3	&	-11 28 08	&	16.6	&	1.6	&	14.9	&	14.3	&	14.1	&	$>$50	&	1.0-1.5	 \\
59	&	30	&	1.9	$\pm$	1.0	&	2.0	&	30.31	&	7 02 51.3	&	-11 24 03	&	17.5	&	4.0	&	14.6	&	13.7	&	13.4	&	05-10	&	1.0-1.5	 \\
60	&	30	&	1.5	$\pm$	0.9	&	1.6	&	30.20	&	7 03 30.4	&	-11 48 07	&	13.7	&	2.4	&	13.5	&	13.0	&	12.8	&	20-50	&	1.0-1.5	 \\
61$^x$	&	30	&	1.7	$\pm$	0.9	&	1.9	&	30.27	&	7 03 33.5	&	-11 34 27	&	15.2	&	2.6	&	13.5	&	12.8	&	12.5	&	05-10	&	1.0-1.5	 \\
\hline																													
62	&	30	&	1.5	$\pm$	0.9	&	1.7	&	30.20	&	7 03 42.1	&	-11 35 14	&	15.5	&	2.5	&	13.4	&	12.6	&	12.4	&	02-05	&	1.5-2.0	 \\
63$^{*,g}$	&	30	&	2.5	$\pm$	1.0	&	2.5	&	30.43	&	7 03 43.2	&	-11 33 06	&	9.1	&	3.4	&	6.5	&	5.2	&	3.8	&	??	&	??	 \\
64$^x$	&	30	&	2.9	$\pm$	1.1	&	2.7	&	30.49	&	7 03 47.5	&	-11 31 49	&	13.8	&	1.3	&	13.5	&	12.8	&	12.6	&	01-02	&	0.5-1.0	 \\
65	&	30	&	1.2	$\pm$	0.8	&	1.5	&	30.11	&	7 03 48.2	&	-11 41 46	&	14.0	&	0.3	&	13.6	&	12.8	&	12.6	&	??	&	??	 \\
66$^{j,x}$	&	30	&	6.2	$\pm$	1.4	&	4.3	&	30.82	&	7 03 51.5	&	-11 34 56	&	--	&	--	&	11.0	&	10.8	&	10.7	&	01-02	&	3.0-5.0	 \\
\hline
\end{tabular}
}
\end{center}
Columns description: (1) Source name: CMaX~1 to 56 (field 1), CMaX~62 to 98 (field 2), CMaX~57 to 61 (both fields);
(2-5) X-ray data; (6-7) Coordinates of the optical counterpart, when available. (8-9) R magnitude and visual extinction ($A_V$), obtained from 
 available catalogues. (10-12) Near-IR magnitudes from the 2MASS Catalogue; (13-14) The estimation of age and mass is based on near-IR. 

Notes: Stars previously identified (Clari\'a  1974, indicated by ``C" and Shevchenko et al. 1999, indicated by ``SEI"):
(a) HD~51896 (F5),	(b)	SS~75, (c) HD~52014 (F5),	(d)	HD~52385 (K0),	(e)	GU~CMa, HD~52721 (B2V), C~63 ,  SEI~158;	
(f)	FZ ~CMa (B2.5), C~67, SEI~159;	(g)	Z~CMa (Bpe), C~76, SEI~161; (h)	 BD~-111762 (B2), C~77, SEI~92; (i) C~14, SEI~94.	
Members of clusters: (j)  vdB-RN92, (k) BRC~27. Objects detected by {\it XMM-Newton} and/or {\it Chandra} are indicated by ``x".
Relevant sources: (*) H-K excess.
\end{table*}

%%-----------------------------Table 1 continued
\begin{table*}[!t]
\caption{Table A1 (continued)}
\smallskip
\begin{center}
{\scriptsize
\begin{tabular}{cccccccccccccc}
\hline
\noalign{\smallskip}
CMaX &	Err&Cnt/ks	&			S/N&		logLx &	R.A.  	&	Dec.  	&	R	&	$A_V$	&	J	  &	H	     &	K	       &	Age	&	Mass	\\
         &    (")      &         &               &             & J2000  &  J2000    & (mag) &  (mag)  &   (mag) &    (mag) &  (mag)   &  (Myr)    &   ($M_{\odot}$)  \\
\noalign{\smallskip}
\hline
67$^x$	&	30	&	2.1	$\pm$	1.0	&	2.2	&	30.35	&	7 03 52.5	&	-11 26 17	&	16.0	&	1.4	&	13.7	&	12.9	&	12.6	&	??	&	??	 \\
68$^j$	&	30	&	2.0	$\pm$	1.0	&	2.1	&	30.34	&	7 03 54.0	&	-11 32 37	&	14.1	&	2.3	&	14.6	&	13.7	&	13.3	&	2-5	&	0.5-1.0	 \\
69$^h$	&	30	&	7.9	$\pm$	1.6	&	4.9	&	30.93	&	7 03 54.4	&	-11 28 29	&	9.4	&	1.5	&	9.4	&	9.4	&	9.4	&	$<$1?	&	$>$5	 \\
70	&	30	&	2.2	$\pm$	1.0	&	2.2	&	30.37	&	7 03 54.8	&	-11 42 37	&	15.1	&	0.3	&	12.0	&	11.5	&	11.3	&	01-02	&	1.5-2.0	 \\
71$^x$	&	30	&	4.9	$\pm$	1.3	&	3.7	&	30.72	&	7 03 55.7	&	-11 29 32	&	14.9	&	2.5	&	12.9	&	12.2	&	12.0	&	02-05	&	1.5-2.0	 \\
72$^{i,j}$	&	30	&	1.6	$\pm$	0.9	&	1.9	&	30.25	&	7 03 56.8	&	-11 34 42	&	11.0	&	7.0	&	10.5	&	10.2	&	10.1	&	??	&	??	 \\
73$^{j,x}$	&	30	&	2.9	$\pm$	1.1	&	2.7	&	30.50	&	7 04 00.4	&	-11 34 00	&	12.7	&	7.0	&	12.1	&	10.9	&	10.2	&	$<$1	&	2.0-3.0	 \\
74$^{*,k}$	&	30	&	9.2	$\pm$	1.8	&	5.3	&	31.00	&	7 04 01.4	&	-11 23 35	&	12.6	&	2.5	&	11.5	&	10.8	&	10.3	&	02-05	&	2.0-3.0	 \\
75$^{k,l,x}$	&	30	&	6.4	$\pm$	1.5	&	4.3	&	30.84	&	7 04 02.3	&	-11 25 39	&	10.8	&	1.3	&	10.4	&	10.3	&	10.3	&	1-2	&	3.0-5.0	 \\
76$^m$	&	30	&	3.8	$\pm$	1.3	&	3.0	&	30.61	&	7 04 06.6	&	-11 18 35	&	10.9	&	6.0	&	13.1	&	12.2	&	11.6	&	02-05	&	2.0-3.0	 \\
77$^m$	&	30	&	3.0	$\pm$	1.1	&	2.7	&	30.51	&	7 04 09.2	&	-11 30 08	&	14.2	&	0.8	&	13.7	&	13.0	&	12.9	&	01-02	&	1.0-1.5	 \\
78	&	30	&	2.2	$\pm$	1.2	&	1.9	&	30.38	&	7 04 11.3	&	-11 16 48	&	13.4	&	3.5	&	12.0	&	11.4	&	11.1	&	05-10	&	2.0-3.0	 \\
79	&	30	&	4.1	$\pm$	1.3	&	3.2	&	30.64	&	7 04 12.1	&	-11 21 10	&	13.8	&	2.9	&	14.4	&	13.4	&	13.1	&	??	&	??	 \\
80	&	30	&	2.1	$\pm$	1.0	&	2.1	&	30.36	&	7 04 12.3	&	-11 39 23	&	12.0	&	1.2	&	11.3	&	11.0	&	11.0	&	05-10	&	2.0-3.0	 \\
81$^{m,n}$	&	30	&	3.2	$\pm$	1.2	&	2.7	&	30.54	&	7 04 13.2	&	-11 19 01	&	7.8	&	3.5	&	10.9	&	10.6	&	10.5	&	01-02	&	3.0-5.0	 \\
82$^{o,x}$	&	30	&	3.8	$\pm$	1.2	&	3.0	&	30.61	&	7 04 15.9	&	-11 24 06	&	9.4	&	1.4	&	9.0	&	8.9	&	8.9	&	$<$1?	&	$>$5	 \\
83$^p$	&	30	&	3.5	$\pm$	0.8	&	4.3	&	30.58	&	7 04 17.4	&	-11 43 10	&	10.9	&	0.6	&	10.2	&	10.1	&	10.0	&	02-05	&	3.0-5.0	 \\
84$^x$	&	30	&	3.5	$\pm$	0.8	&	4.3	&	30.58	&	7 04 18.3	&	-11 42 36	&	13.8	&	0.0	&	12.0	&	11.5	&	11.3	&	02-05	&	1.5-2.0	 \\
85	&	30	&	4.2	$\pm$	1.3	&	3.3	&	30.66	&	7 04 18.8	&	-11 24 47	&	14.7	&	0.8	&	14.6	&	13.9	&	13.7	&	05-10	&	0.5-1.0	 \\
86$^x$	&	30	&	2.9	$\pm$	1.1	&	2.6	&	30.50	&	7 04 19.1	&	-11 33 48	&	13.4	&	0.4	&	13.1	&	12.4	&	12.2	&	02-05	&	0.5-1.0	 \\
87	&	30	&	3.0	$\pm$	1.2	&	2.6	&	30.51	&	7 04 20.0	&	-11 22 22	&	13.2	&	0.9	&	14.4	&	13.3	&	12.7	&	??	&	??	 \\
88	&	30	&	1.8	$\pm$	0.9	&	1.9	&	30.29	&	7 04 20.6	&	-11 36 44	&	16.7	&	0.7	&	14.6	&	13.7	&	13.5	&	??	&	??	 \\
89$^x$	&	30	&	3.5	$\pm$	1.2	&	2.9	&	30.57	&	7 04 26.3	&	-11 31 21	&	13.8	&	0.8	&	12.5	&	11.6	&	11.2	&	??	&	??	 \\
90	&	30	&	4.0	$\pm$	1.3	&	3.1	&	30.63	&	7 04 25.3	&	-11 34 28	&	13.2	&	1.1	&	11.7	&	11.1	&	10.9	&	$<$1	&	1.5-2.0	 \\
91$^x$	&	30	&	5.2	$\pm$	1.7	&	3.1	&	30.75	&	7 04 29.8	&	-11 47 21	&	12.1	&	0.0	&	11.2	&	10.5	&	10.3	&	??	&	??	 \\
92$^x$	&	30	&	2.3	$\pm$	1.1	&	2.2	&	30.39	&	7 04 31.0	&	-11 32 42	&	14.0	&	0.8	&	13.9	&	13.1	&	12.9	&	01-02	&	0.1-0.5	 \\
93	&	30	&	2.8	$\pm$	1.1	&	2.4	&	30.47	&	7 04 37.0	&	-11 30 59	&	14.0	&	0.0	&	14.0	&	13.8	&	13.7	&	??	&	??	 \\
94	&	30	&	1.6	$\pm$	1.0	&	1.5	&	30.23	&	7 04 48.4	&	-11 15 07	&	15.7	&	0.4	&	14.8	&	14.4	&	14.1	&	$>$50	&	1.0-1.5	 \\
95	&	30	&	5.4	$\pm$	1.8	&	3.0	&	30.76	&	7 04 52.5	&	-11 38 21	&	14.7	&	0.2	&	13.2	&	12.4	&	12.1	&	??	&	??	 \\
96$^x$	&	30	&	3.4	$\pm$	1.5	&	2.3	&	30.56	&	7 04 56.3	&	-11 29 33	&	12.5	&	1.2	&	11.5	&	10.9	&	10.2	&	01-02	&	2.0-3.0	 \\
97	&	30	&	0.9	$\pm$	0.6	&	1.7	&	30.00	&	7 05 15.5	&	-11 31 06	&	14.8	&	0.0	&	13.2	&	12.8	&	12.7	&	20-50	&	1.0-1.5	 \\
98	&	30	&	0.9	$\pm$	0.6	&	1.7	&	30.00	&	7 05 15.7	&	-11 32 05	&	14.4	&	0.0	&	12.8	&	12.1	&	12.0	&	01-02	&	1.0-1.5	 \\
\noalign{\smallskip}
\hline
\end{tabular}
}
\end{center}
Notes: Stars previously identified (Clari\'a  1974, indicated by ``C" and Shevchenko et al. 1999, indicated by ``SEI"):
 (l) SEI~99;  (n) SEI~111; (o) HD~53339 (B3), C84, SEI114; (p) HD~53396  (B9), SEI~112.
Members of clusters: 	(k) BRC~27; (m) NGC~2327. Objects detected by {\it XMM-Newton} and/or {\it Chandra} are indicated by ``x" (see list in Table A3).
\end{table*}

%%-------------------------------------------------- Figure A1
\begin{figure*}[]
\vskip 0.2cm
\hskip 12.6cm
\includegraphics[height=5cm,angle=270]{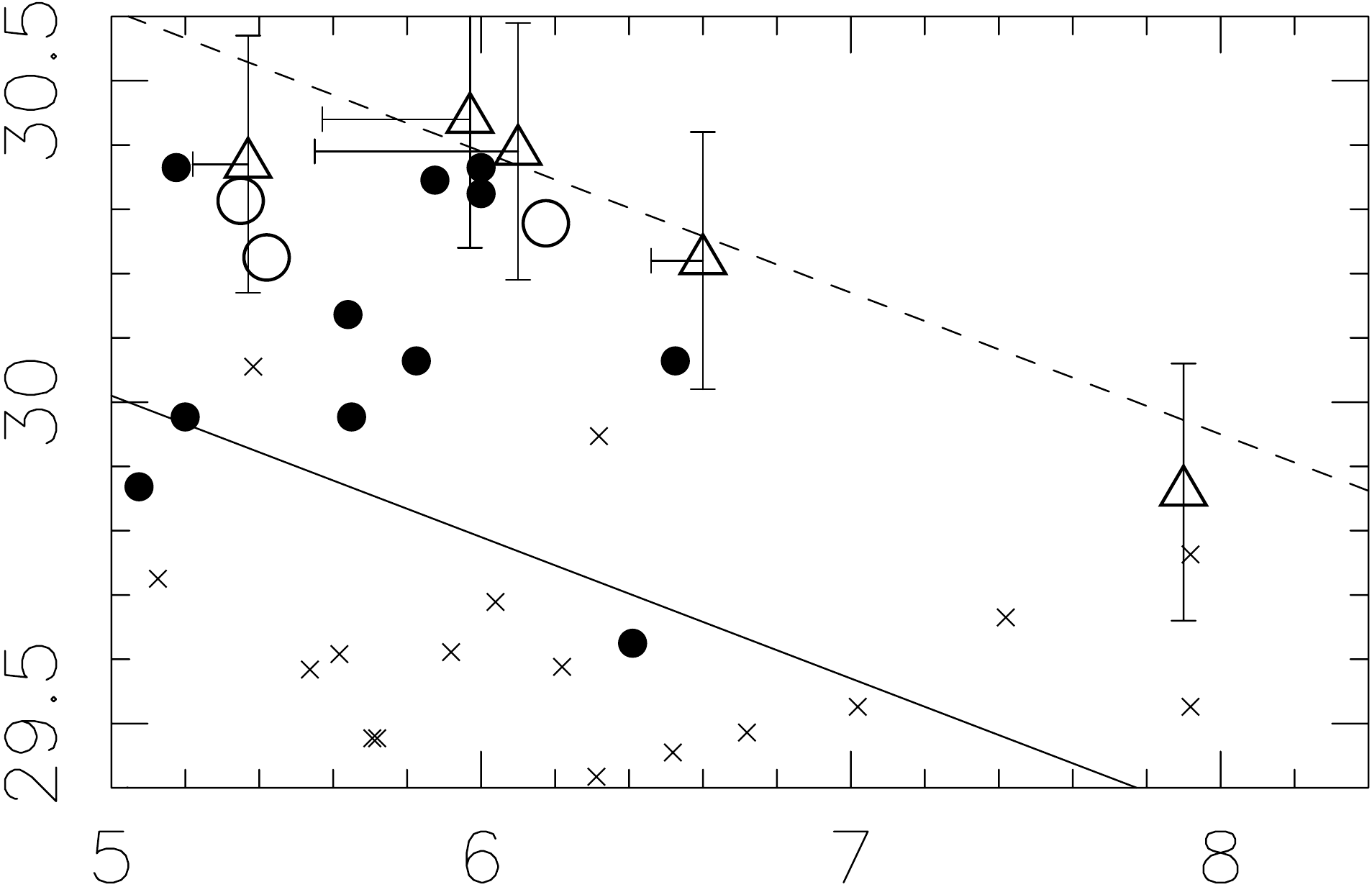}
\vskip -3.5cm
\includegraphics[height=18cm,angle=270]{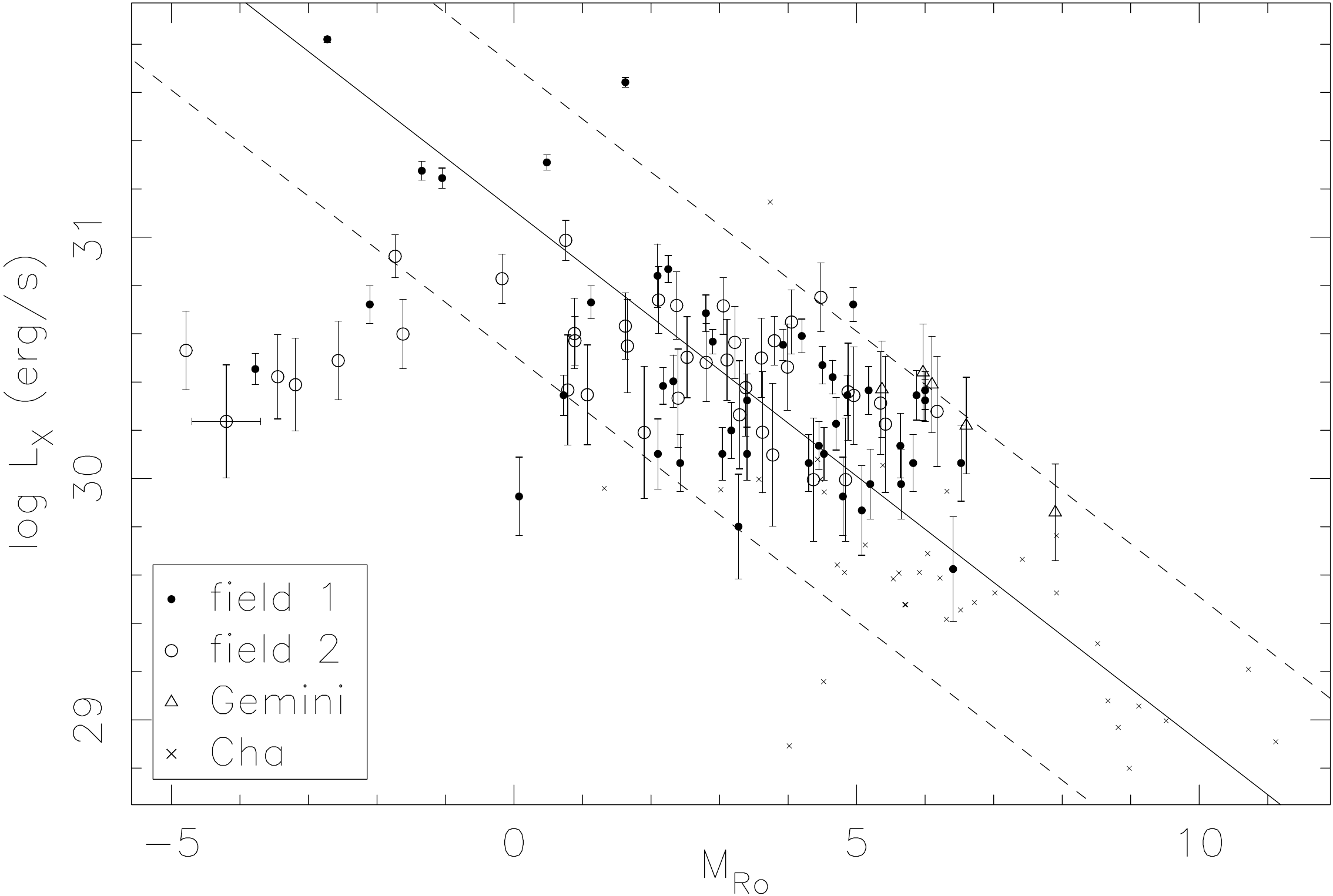}
\caption{The distribution of the CMaX sources (circles)
in the diagram of log($L_X$) as a function of de-reddened absolute magnitudes 
$M_{Ro}$. Error bars of the optical magnitudes are shown for two illustrative examples.
Triangles indicate the five counterparts detected by {\it Gemini}, for which details are shown in the corner box.
In this case, the horizontal bars show the possible range of magnitudes, when one or all counterparts
contribute to the integrated X-ray luminosity (see text, Sect. 4.4).
The Cha I X-ray data, represented by crosses, were extracted from Feigelson 
et al. (1993). The full line indicates the correlation obtained for the Cha I sources, 
which  have well determined $L_X$ and extinction. Dashed lines are used to indicate 
the 2$\sigma$ deviation from the linear fit. }
\end{figure*}

%%-------------------------------------------------- Figure A2
\begin{figure*}[]
\vskip 0.2cm
\hskip 12.6cm
%\vskip 1.5cm
\includegraphics[height=5cm,angle=270]{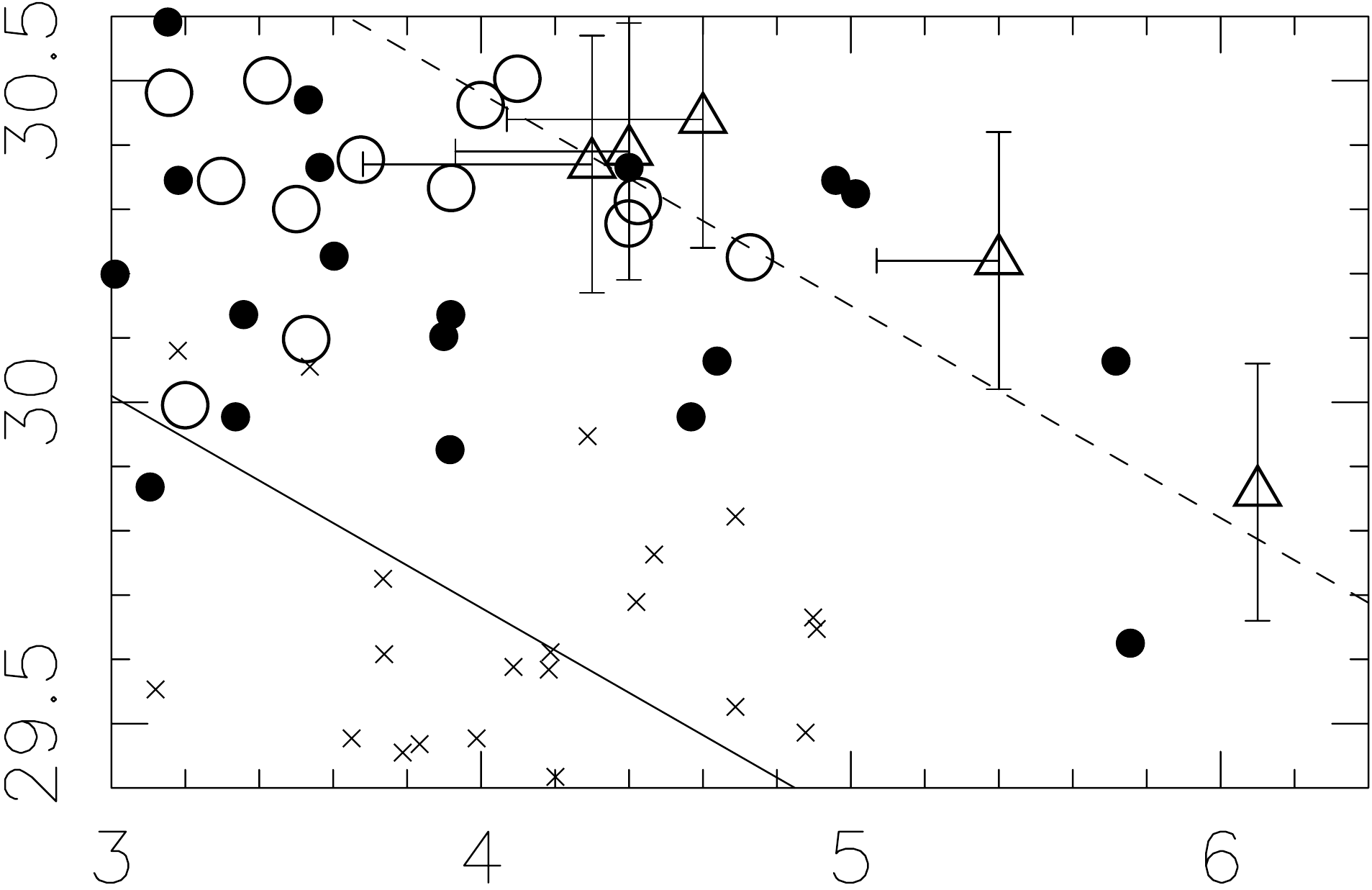}
\vskip -3.5cm
\includegraphics[height=18cm,angle=270]{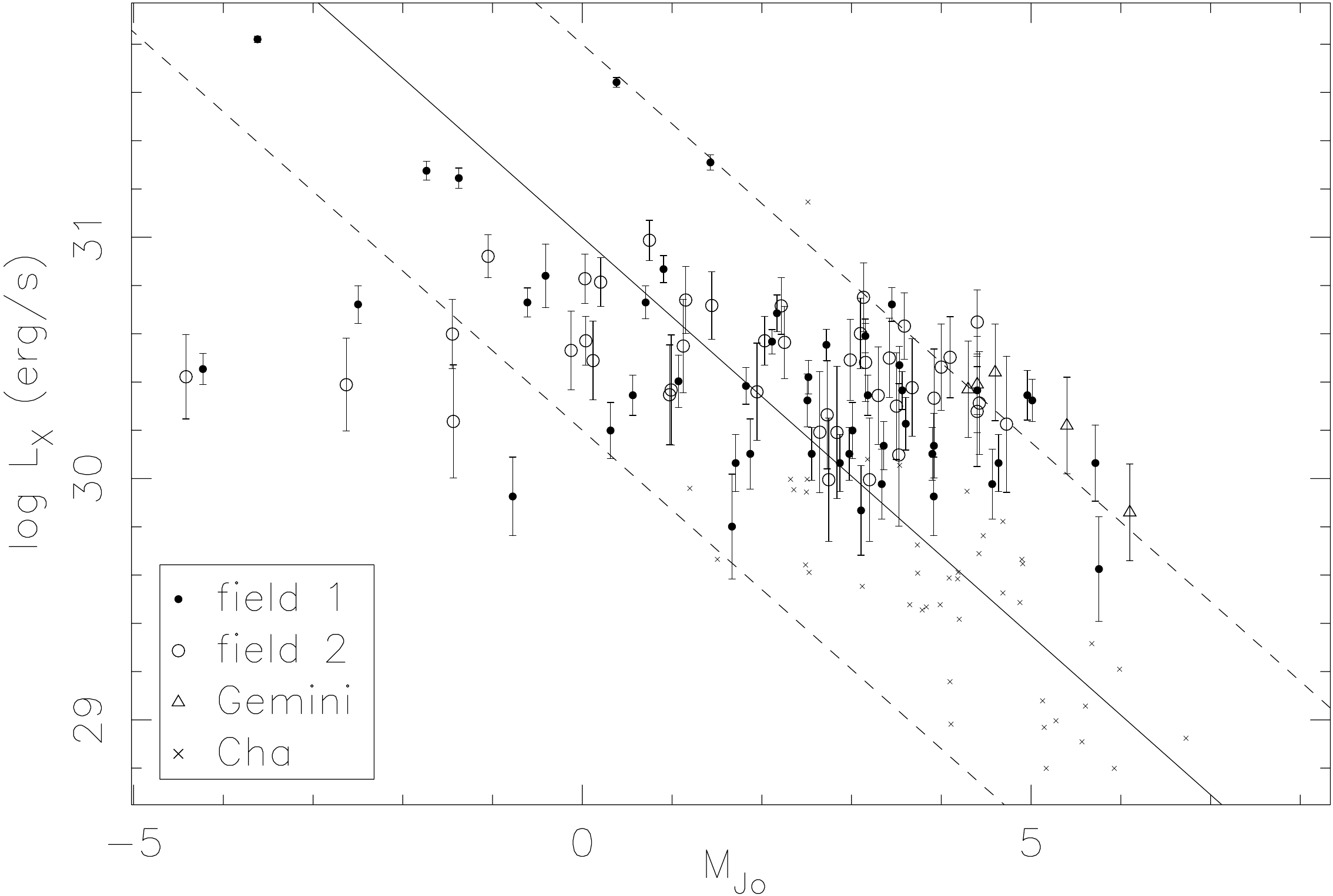}
\caption{The same as Fig. A.1, showing the diagram of log($L_X$) as a function of de-reddened absolute magnitudes 
$M_{Jo}$. Error bars for J magnitudes are not plotted, since they are smaller than the symbols for most of the sources.}
\end{figure*}

%%--------------------------------------------------------------sect. A.2
\subsection{X-ray luminosity compared to optical and near-IR luminosities}

Figure A1 shows the diagrams $L_X$ versus $M_{Ro}$ (dereddened absolute magnitude) of the sample,  compared to the correlation of X-ray emission 
with the  optical absolute magnitude, expected for young stars. 
This correlation was obtained by comparing the magnitudes to the mean X-ray luminosities (\~L$_X$) 
estimated  for {\it ROSAT} sources detected in the Chamaeleon I (Cha  I) cloud, adopting the conversion
1 cnt~ks$^{-1}$= $3 \times 10^{28}$ erg~s$^{-1}$ (Feigelson et al.  1993).

Taking into account that the extinction cross-section at 2 $\mu$m and at 2 keV are almost the same (e.g., Ryter 1996), the X-ray emission was 
compared to the J band  magnitude in order to check whether these data show the same correlation as previously found for young, low-mass stars, i.e.,
the log(\~L$_X$) versus $M_{Jo}$ diagram  (Casanova et al. 1995), which is  presented in
 Figure A2. The Cha I sources were also plotted for comparison: since they are much closer ($d = 140$ pc) than CMa R1, the same {\it ROSAT} sensitivity 
allows to probe fainter, lower-mass young stars.

The absolute magnitudes of the Cha I sources 
were estimated by adopting a distance modulus of 5.73 and were 
dereddened by using their visual extinctions, when available. Otherwise, A$_V$=1mag was adopted.
The Cha I correlation was established on the basis of 16 sources with well-determined magnitudes
and extinctions (see Casanova et al. 1995 for details; also Lawson et al. 1996). 
Figure A1 shows the distribution of the Cha I sources and the 
corresponding linear regression: 
log(\~L$_X$[erg~s$^{-1}$]) = 31.10($\pm$0.3) - 0.22$M_{Ro}$, and for  
the J band log(\~L$_X$[erg~s$^{-1}$]) = 31.00($\pm$0.4) - 0.33$M_{Jo}$. Dashed lines are used to indicate
the 2$\sigma$  deviation.

Both diagrams, log(\~L$_X$)  vs. $M_{Jo}$, and log(\~L$_X$)  vs. $M_{Ro}$, show  that total absolute magnitudes are compatible 
with the X-ray luminosities typical of young stars, but tend to deviate from the Cha I correlation as the counterparts become brighter. Also, in 
the upper right-hand box of the diagram in Figure A1, we zoom on the {\it Gemini} counterparts. For these sources, the {\it Gemini} points are 
above the correlation, indicating that other, undetected (absorbed ?) stars may also contribute to the unresolved X-ray emission. We defer a 
more detailed analysis of these diagrams to a later study.

%%-----------------------------Table A3
\begin{table}[]
\caption{List of {\it ROSAT} sources observed by {\it XMM-Newton} or {\it Chandra}}
\smallskip
\begin{center}
{\scriptsize
\begin{tabular}{ccccc}
\hline
\noalign{\smallskip}
CMaX &	2MASS &	Source$^a$ & R.A.$^b$	&	Dec.$^b$	\\
\noalign{\smallskip}
\hline
61	&	07033347-1134269	&	C20	&	07 03 33.46	&	-11 34 26.7	\\
64	&	07034751-1131489	&	C47	&	07 03 47.50	&	-11 31 48.6	\\
66	&	07035152-1134557	&	C54	&	07 03 51.50	&	-11 34 55.7	\\
67	&	07035249-1126168	&	C57	&	07 03 52.38	&	-11 26 17.7	\\
71	&	07035575-1129315	&	C66	&	07 03 55.73	&	-11 29 31.3	\\
73	&	07040041-1133596	&	C77	&	07 04 00.35	&	-11 33 59.2	\\
75	&	07040234-1125393	&	X41	&	07 04 02.48	&	-11 25 37.5	\\
82	&	07041588-1124055	&	X54	&	07 04 15.94	&	-11 24 04.0	\\
84	&	07041833-1142359	&	X09	&	07 04 18.42	&	-11 42 33.9	\\
86	&	07041912-1133480	&	X14	&	07 04 19.21	&	-11 33 48.8	\\
89	&	07042625-1131207	&	X11	&	07 04 26.35	&	-11 31 20.0	\\
91	&	07042982-1147208	&	X06	&	07 04 29.97	&	-11 47 19.9	\\
92	&	07043099-1132417	&	X32	&	07 04 31.11	&	-11 32 38.9	\\
96	&	07045632-1129332	&	X13	&	07 04 56.44	&	-11 29 32.0	\\
\noalign{\smallskip}
\hline
\end{tabular}
}
\end{center}
Note: (a) Source number in the {\it Chandra} (C) or {\it XMM-Newton} (X) images;
(b) respective J2000 coordinates. Positional deviations from the coordinates
of near-IR counterparts ({\it 2MASS})
are less than 2 arcsec for {\it Chandra} sources and less than 3.3 arcsec for {\it XMM-Newton}
sources.
\end{table}

%%========================================Appendix B
\section{Results of the Gemini observations}

In this Appendix we show the continuation of the Figures presented in the body of the paper. 

\begin{figure*}[ht]
\begin{center}
\includegraphics[width=5.5cm,clip]{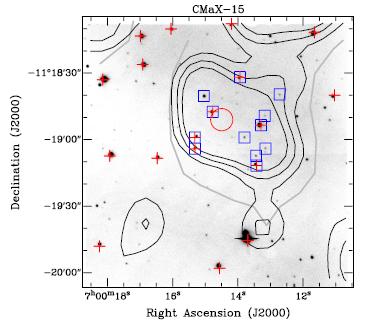}
\includegraphics[width=5.5cm,clip]{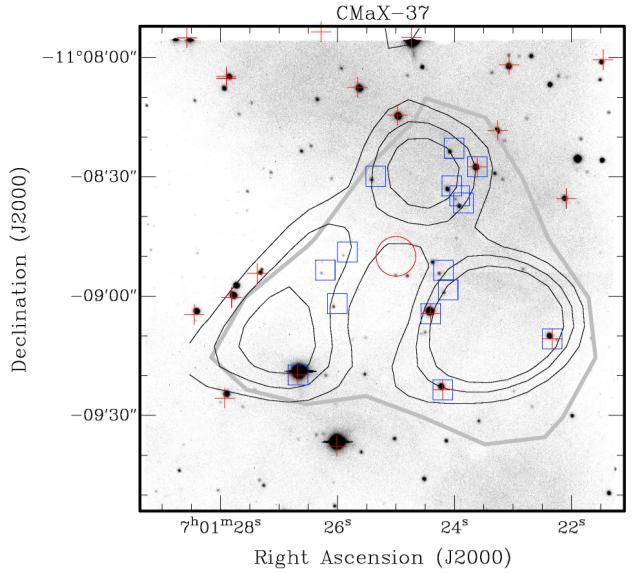}
\includegraphics[width=5.5cm,clip]{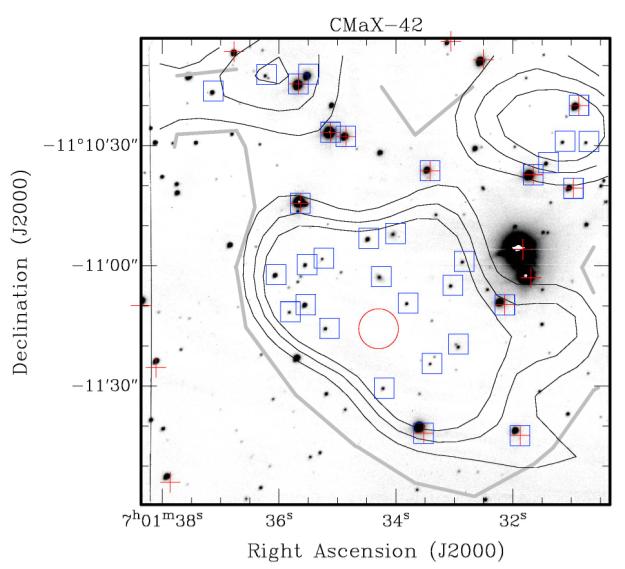}
\caption{Optical image (Gemini I band) of CMaX-15 (left), 37 (middle), and 42 (right). The same as Figure 6.} 
\end{center}
\end{figure*}

%-------------------Figure B2 ---------------------------------- Cores Gemini continued

\begin{figure*}[hb]
\includegraphics[height=8cm,angle=0]{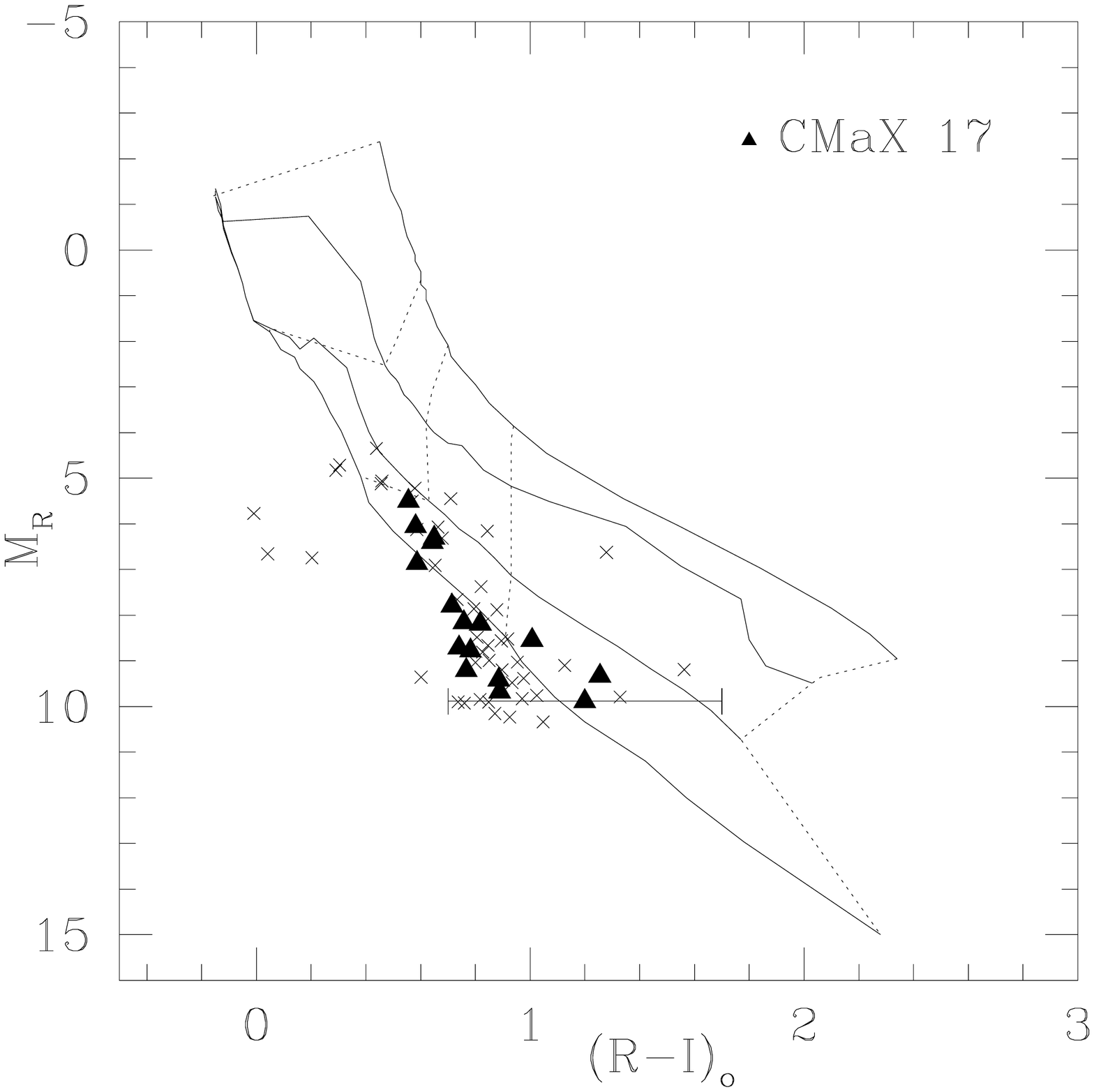}
\includegraphics[height=8cm,angle=0]{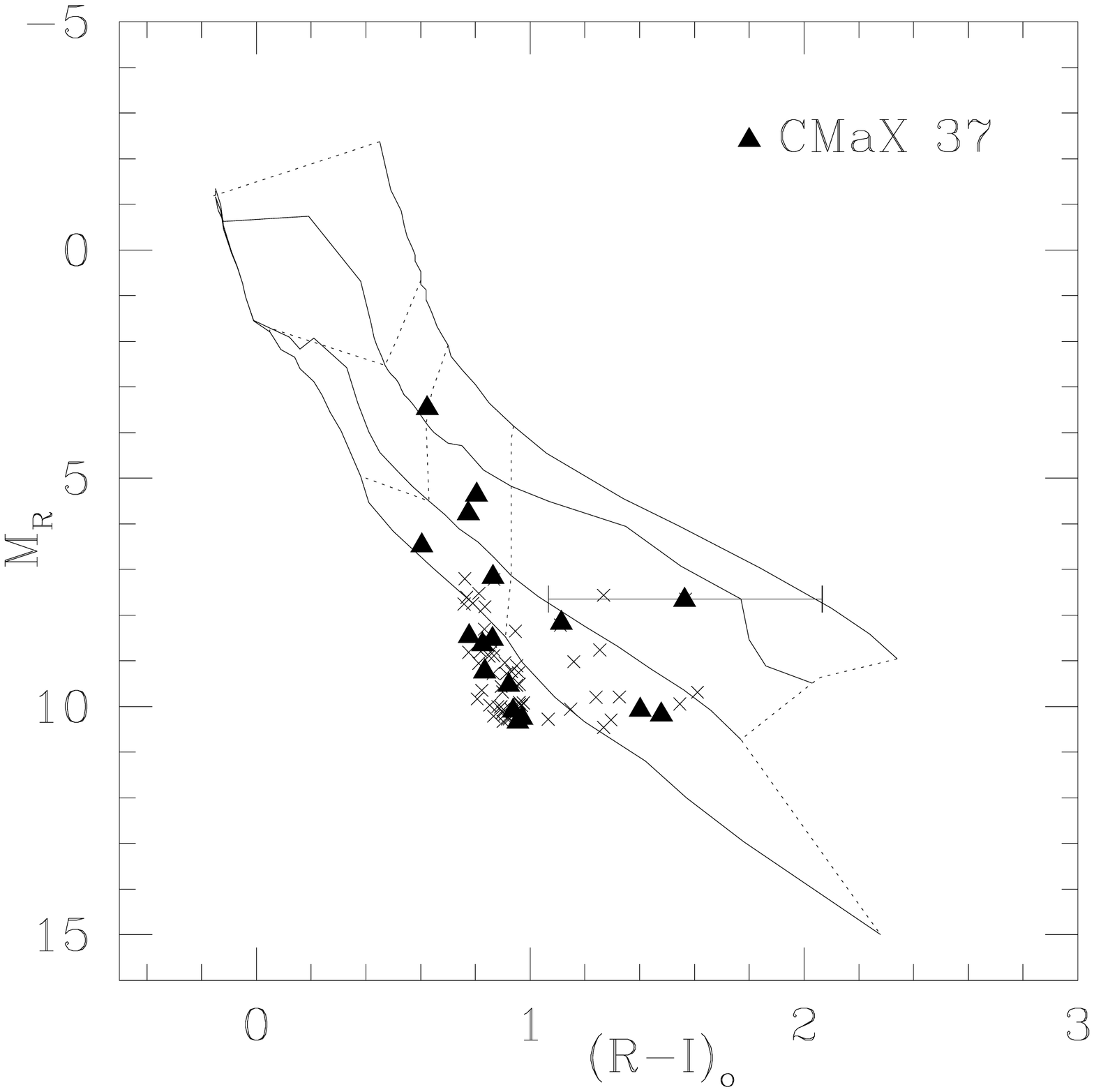}
\includegraphics[height=8cm,angle=0]{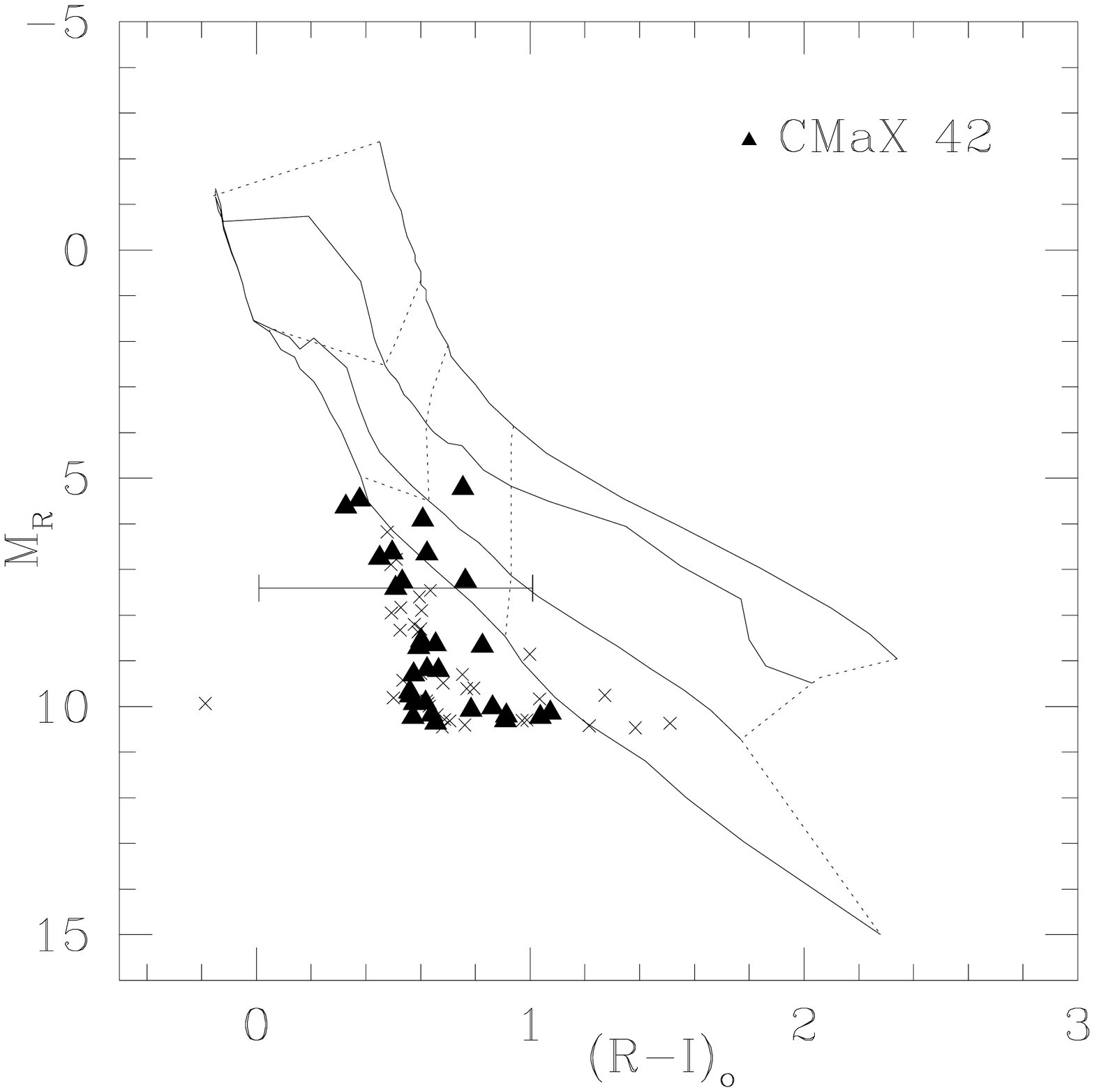}
\hskip 3.3cm
\includegraphics[height=8cm,angle=0]{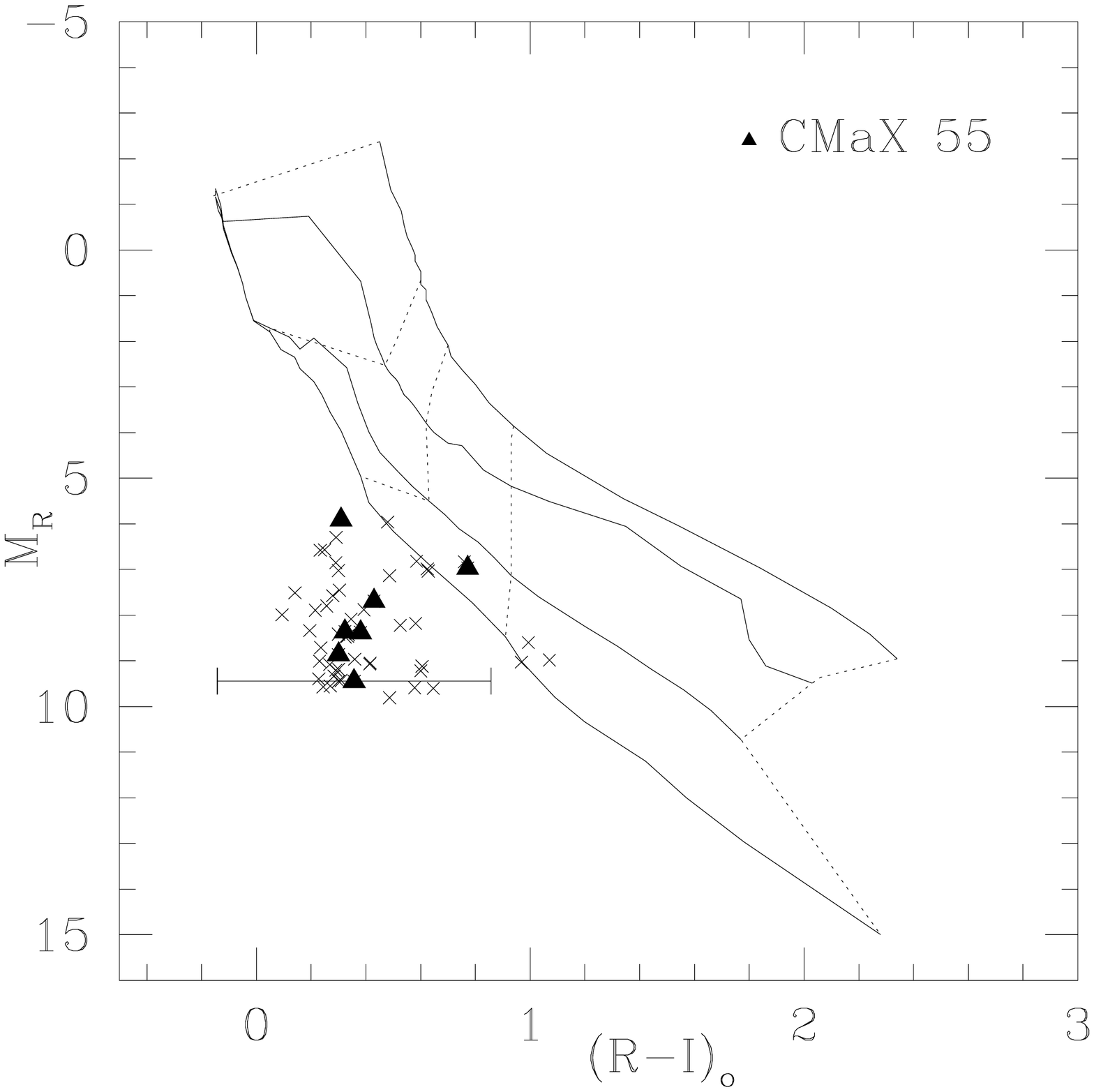}

\caption{Colour-magnitude diagram of the optical candidates to be counterparts of the 
X-ray sources CMaX-17, 37, 42, 55. The same as Figure 8.}
\end{figure*}

\end{appendix}

\end{document}